\documentclass[12pt,preprint]{aastex}

\newcommand{\kms}{km~s$^{-1}$}
\newcommand{\helium}{\ion{He}{1}~$\lambda 10830$}
\newcommand{\heopt}{\ion{He}{1}~$\lambda 5876$}
\newcommand{\pgamma}{P$\gamma$}
\newcommand{\halpha}{H$\alpha$}

\newcommand{\msunyr}{M$_\sun$~yr$^{-1}$}
\newcommand{\vbhi}{$V_{\rm{BHI}}$}
\newcommand{\vc}{$V_c$}
\newcommand{\nod}{\nodata}
\defcitealias{BEK}{BEK}
\defcitealias{HEG}{HEG}

\begin{document}
\slugcomment{Accepted by Astrophysical Journal}
\title{Probing T Tauri Accretion and Outflow with 1 Micron Spectroscopy}

\author{ Suzan Edwards \altaffilmark{1,3,4}, William Fischer \altaffilmark{2},
Lynne Hillenbrand \altaffilmark{3,5}, John Kwan \altaffilmark{2}}

\altaffiltext{1}{Five College Astronomy, Smith College,
 Northampton, MA 10163, sedwards@smith.edu}
\altaffiltext{2} {Five College Astronomy, University of
 Massachusetts, Amherst, MA, 01003,
 kwan@astro.umass.edu, wfischer@astro.umass.edu}
\altaffiltext{3}{Visiting Astronomer, Keck Observatory}
\altaffiltext{4}{Guest Observer, Gemini/Keck-NIRSPEC program}
\altaffiltext{5} {Dept. of Astronomy, California
Institute of Technology, Pasadena, CA 91125,
lah@astro.caltech.edu}

\begin{abstract}

In a high dispersion 1~\micron\ survey of 39 classical T Tauri stars veiling is
detected in 80\%, and \helium\ and \pgamma\ line emission
in 97\% of the stars. On average, the 1~\micron\ veiling
exceeds the level expected from previously identified sources of
excess emission, suggesting the presence of an additional
contributor to accretion luminosity in the star-disk interface
region. Strengths of both lines correlate with veiling, and at
\pgamma\ there is a systematic progression in profile morphology
with veiling.  \helium\ has an unprecedented sensitivity to inner
winds, showing blueshifted absorption below the continuum
in 71\% of the CTTS compared to 0\% at \pgamma.
This line is also sensitive to magnetospheric
accretion flows, with redshifted absorption below the
continuum found in 47\% of the CTTS compared to
24\% at \pgamma.

The blueshifted absorption at \helium\ shows considerable
diversity in its breadth and penetration depth into the
continuum, indicating that a range of inner wind conditions exist in
accreting stars. We interpret the broadest and deepest
blue absorptions as formed from scattering of the
1~\micron\ continuum by outflowing gas whose full acceleration
region envelopes the star, suggesting radial outflow from the
star. In contrast, narrow blue absorption with a range
of radial velocities more likely arises via scattering of the
1~\micron\ continuum by a wind emerging from the inner disk.
Both stellar and disk winds are accretion powered since neither
is seen in non-accreting WTTS and among the CTTS helium strength
correlates with veiling.

\end{abstract}


\section{INTRODUCTION}

In the formation of a star and its attendant planetary system the
phase of disk accretion is always accompanied by mass outflow.
Outflow collimation perpendicular to the disk plane and
correlations between disk accretion rate and wind mass loss
rate, with a rough proportion of ${\dot M}_w/{\dot M}_{acc} \sim
0.1$, certify the intimate relation between these two phenomena
\citep{rich00,cab90}.
Although the coupling of accretion and outflow is well
documented, the means of launching the outflow remains unknown
and is one of the major unsolved mysteries of star formation.
There is a strong motivation to understand where and how
energetic winds originate in accretion disk systems since they
are strong contenders for a major role in mass and angular
momentum transport in the disk  \citep{fc04}, for
stellar spin-down \citep{k91,shu94,mat05}, for heating
the disk atmosphere \citep{glass04}, for
hastening disk dissipation \citep{h00}, and for
disrupting infalling material from the collapsing molecular
cloud core  \citep{t84}.

Most outflow models tap magneto-centrifugal ejection as the
heart of the launching mechanism, where rotating fields in the
disk fling material along inclined field lines and transport
angular momentum away from the disk. A distinguishing
characteristic among disk wind models is the location in the
system where mass loading onto field lines occurs -- either from
the inner disk over a range of radii \citep{kp00}
or from the radius at which the stellar magnetosphere truncates
the disk, lifting most of the accreting material toward the star
into magnetic funnel flows and ejecting the rest along opened
stellar field lines emerging from this point \citep{shu94}.
Others have explored the possibility that accretion-driven winds
emerge via magnetohydrodynamic acceleration from the star rather
than, or in addition to, the disk \citep{hirose,kt88,mat05,love05,rb05}.
Establishing the launch region for accretion powered winds is thus
of considerable interest, since this can discriminate among the various
wind theories and elucidate how angular momentum is removed from
the star-disk environment.

Knowledge of accretion driven winds comes primarily from
spatially extended collimated atomic and molecular jets
comprised of high velocity ($200-400$ \kms) outflowing material
undergoing internal shocks \citep{HEG, reip01,hep04}.
These tracers yield basic wind energetics and establish
the fundamental connection between ${\dot M}_w$ and ${\dot
M}_{acc}$ but provide only indirect information on the wind
launching region. An intriguing result comes from spectroscopy at high
spatial resolution of T Tauri "micro jets" revealing velocity
asymmetries that have been interpreted as coherent rotation of the jet
as would be imparted from an MHD disk wind launched from moderate
disk radii of 0.5-2 AU \citep{cof04}. However, recent three-
dimensional simulations of line profiles formed in precessing jets
with variable ejection velocities demonstrate that these conditions
can produce similar velocity gradients, thereby calling into question
the implied support for magnetocentrifugal wind acceleration models
\citep{cerq06}.

Direct probes of the launch region can, however, only be made via
high resolution spectroscopy. Confoundingly, most
spectral features from sub-AU regions have kinematic signatures
deriving from rotation or accretion rather than outflow. Near
infrared lines of CO and $H_2O$ are modeled as arising from
inner accretion disks in Keplerian rotation at radii from 0.1 to
2 AU, precisely the region where the rotating jets are predicted
to arise \citep{naj03,carr04}. Although modeling these features requires
turbulent broadening and warm disk chromospheres, their symmetry
reveals no evidence for outflowing gas from the upper disk
layers. A possible signature of a wind emerging from the disk
over radii ranging to tens of AU, is seen in the `low velocity
component' (LVC) of cTTS forbidden lines, distinguished from the
`high velocity component' (HVC) by smaller blueshifts ($\sim 10$
\kms), smaller spatial extensions, and line ratios indicative of
formation under conditions of higher density, although its
interpretation as the launch of an extended disk wind remains
unconfirmed \citep {HEG,kt95}.

However, outflow velocities of 200 to 500 \kms\ characteristic of
jets and HVC forbidden line emission point to an origin deeper
in the potential well of the accretion disk system -- either the
very inner disk, the star-disk interface, or the star itself.
These regions, where the stellar magnetosphere controls the
accretion flow onto the star, are probed via permitted atomic
emission lines from the infrared through to the ultraviolet and
include a wide range of ionization states. Their line profiles,
typically with strong and broad emission sometimes accompanied
by redshifted absorption below the continuum and/or narrow
centered emission, are attributed to formation in magnetospheric
funnel flows and accretion shocks on the stellar surface
\citep{e94,mch01,her02,her05}. Until recently the only clear
evidence for inner winds was blueshifted absorption superposed
on the broad emission lines of H$\alpha$, Na D, Ca II H\&K, and
MgII h\&k , signifying the presence of a high velocity wind
close to the star \citep{cal97,ard02} but yielding little
information on the nature or location of the wind launching region.

Somewhat surprisingly, robust diagnostics of the inner wind in
accretion disk systems are turning out to be lines formed under
conditions of high excitation -- He I and possibly also C III and O VI
\citep{e03,dup05} -- arising
in regions where temperatures exceed 15,000 K.
The first indication of inner winds from regions with high
excitation came from a study of He I
$\lambda$5876 and $\lambda$6678 in a sample of 31 accreting cTTS.
\citep[hereafter BEK] {BEK}. The high excitation of the helium lines
 restricts line formation to a region either of high temperature or close
proximity to a source of ionizing radiation. In spite of these
restrictions, the helium emission lines were found to have a
composite origin, including contributions from a wind, from the
funnel flow, and from an accretion shock.

Confirmation that broad helium emission traces inner winds in
accreting stars came from observations of another line in the He
I triplet series, $\lambda$10830, which is a far more sensitive
diagnostic of outflowing gas. This line, immediately following
$\lambda$5876 in a recombination-cascade sequence, has a highly
metastable lower level, setting up a favorable situation for
tracing outflowing gas in absorption. An initial look at 6 high
accretion rate stars (AS 353A, DG Tau, DL Tau, DR Tau, HL Tau
and SVS 13) focused on examples of \helium\ profiles with
exceptionally strong P Cygni wind signatures \citep{e03}. More recently
\helium\ profiles for two additional cTTS (TW Hya and T Tau)
were also shown to have P Cygni profiles \citep{dup05}. Both
of these studies concluded that the winds tracked by helium
absorption likely arise as a wind expanding radially away from
the star, tracing the wind through its full acceleration from
rest to several hundred \kms.

In this manuscript we present results of a census of
He I $\lambda$10830 in 39 cTTS and 6 wTTS plus 3 Class I sources
from spectra collected with Keck II's NIRSPEC. We find that
this line is exceptionally sensitive not only to winds but also to
infalling gas. Our spectra also yield simultaneous
\pgamma\ profiles and 1~\micron\ continuum excess (veiling),
providing a good combination of features to probe the relation
between inner winds and accretion in young low mass stars. The
remainder of the paper is organized as follows: {\it Section 2:}
observations and data reduction, including sample selection and
a description of the determination of the excess continuum
emission and residual emission profiles, {\it Section 3:}
accretion diagnostics: \pgamma\ and 1~\micron\ veiling, {\it
Section 4:} wind diagnostic: \helium\, {\it Section 5:} Class I
sources, {\it Section 6:} discussion of magnetospheric accretion
contributions to \pgamma\ and \helium\ and wind contributions to
\helium, {\it Section 7;} origin of 1~\micron\ veiling, {\it
Section 8:} conclusions.

\section{OBSERVATIONS AND DATA REDUCTION}

Near-infrared echelle spectra of 39 classical T Tauri stars
(CTTS), 6 weak T Tauri stars (WTTS) and 3 Class I sources were
acquired with NIRSPEC on Keck II \citep{mclean98}, using the N1
filter ($Y$ band) with wavelength coverage from 0.95 to
1.12~\micron\ at $R=25,000$ ($\Delta V=12$~\kms). The detector
is a 1024 x 1024-pixel InSb chip, and the spectrograph slit was
0.43'' x 12.'' Spectra were taken in nodded pairs in an ABBA
pattern, with the target separated by 5" in each AB pair.
Typical exposure times were 300 seconds, integrated in each
case to acquire S/N $\sim$ 70-80 in the reduced 1-D spectra.
All objects were observed on a 3 night run in
November 2002 by Hillenbrand and Edwards. Three of our program
stars were also observed on 22 November 2001 for Edwards and A.
Dupree by the Gemini-Keck queue observers T. Geballe and M.
Takamiya (see \citealt{e03}). In 2002 spectra were acquired
without the image de-rotator, due to equipment malfunction.
Although this prevents study of spatially extended emission, as
reported for DG Tau \citep{tak02}, it is not a problem for
extraction of a spectrum from the central point source, where
most of the line emission and all of the line absorption arises.

Spectra were processed in IRAF using a modified version of a
script originally provided by M. Takamiya. Data reduction steps
included spatial rectification of the 2D images, wavelength
calibration with a Xe/Ar/Kr/Ne arc lamp spectrum, correction for
telluric emission, telluric absorption, and instrumental
fringing, and extraction to 1-D spectra using IRAF's {\it apall}
package. Telluric emission was removed by subtraction of each AB
pair of images, yielding two sky-subtracted
spectra which were then coadded after spatial rectification and
wavelength calibration. Telluric absorption features were
removed by dividing the rectified, extracted, co-added,
wavelength-calibrated spectra by the spectrum of an early-type
standard observed at a similar airmass.

Division by a telluric standard also corrected the sinusoidal
fringing pattern  characterized by a spacing of $\sim 1.5$~\AA\
(10 pixels) and an amplitude $\sim 3\%$ of the continuum)
for the 2002 data, where there was a
good alignment ($\le 0.7$ pixels) of the fringing
pattern between the target and telluric standard. For the few
2001 spectra taken under queue observing, the fringe pattern was
misaligned between target and standard by 1.5 to 4.5 pixels (0.5
to 1.5 times the instrumental resolution). For these spectra we
determined the amount of fringe shift in wavelength by
cross-correlating the target and standard spectra in a region
free of spectral features, and found that the telluric
absorption features shared the same misalignment. Consequently,
after shifting the standard by the requisite amount, division by
the standard conveniently removed both contaminating features. A
final problem occasionally arose for stars observed at high air
mass, when weak residual telluric features in the vicinity of
the \helium\ and \pgamma\ lines appeared in the reduced
spectrum. These were readily removed from the residual profile
by determining their strength relative to stronger telluric
features in the rest of the spectral order.

Our telluric standards were of early spectral type (O7 to B9),
providing a nearly featureless continuua over most of the orders
yielding a good definition of telluric features. However, they
have strong photospheric lines of \helium\ and \pgamma, our two
primary spectral diagnostics. We thus took special care to
define and remove these broad photospheric features in the
standards so that spurious emission would not be induced in our
T Tauri spectra. This was accomplished by first creating a
synthetic spectrum of each standard where the observed profiles
were modeled with a combination of Gaussian and Lorentzian
functions, and then dividing the synthetic spectrum into the
actual standard spectrum. We judged the fit effective when the
final continuum-normalized spectra were flat in these spectral
regions to within one percent. An additional correction for
helium emission had to be made for HR 130 using a similar
technique. The absence of any broad helium or hydrogen emission
in our WTTS spectra (discussed in the following section)
verifies that this procedure was effective. After removal of all
of the above instrumental effects, the definition of the final
spectrum is limited only by the signal to noise of the target and
standard.

Although 13 echelle orders appear on the chip with the N1 filter,
the first and last are severely contaminated by telluric
absorption. Of the remaining eleven orders, the number and
location of arc lamp lines is such that reliable wavelength
solutions can be obtained for four full orders and five partial
ones. The order extending from 1.081 to 1.096~\micron\ contains
both our prime spectroscopic diagnostics of \helium\ and
\pgamma\ and also many photospheric features in late type stars.
In addition to this order we identified two additional orders
with abundant photospheric lines that we used for redundant
estimates of the continuum veiling and determination of the
stellar velocity. One, immediately to the blue
of our prime order, extends from 1.066 to 1.081~\micron\ and has
numerous photospheric features and minimal contamination from
telluric absorption. A second, although its full extension is
from 0.971 to 0.986~\micron, only had sufficient arc line
coverage for wavelength calibration from 0.971 to 0.979~\micron.
Even with only half the order usable we found it to be
preferable to other orders because its strong photospheric
features were still detectable when photospheric lines in
the other orders were obscured by broad metallic
emission features which were sometimes found in stars with the
highest continuum veiling.

\subsection{Determination of Excess Continuum Emission and Residual Profiles}

The 39 CTTS and 6 WTTS in our sample are identified in
Table~\ref{t.sample}. All but two have spectral types from K0 to
M5, with K7/M0 stars comprising the majority (27 stars). The
remaining two are of solar type, SU Aur (G2) and GW Ori (G5).
Most (30) are in the Tau-Auriga star formation complex, the rest
are well studied T Tauri stars from associations including those
in Orion and TW Hydra. The sample covers the full range of
continuum emission excesses in the optical and ultraviolet found
for CTTS and was selected to overlap extensively with an earlier
optical echelle survey of 31 CTTS appearing in \citealt{HEG}
(hereafter \citetalias{HEG}) and \citealt{BEK} (hereafter
\citetalias{BEK}), thereby allowing a comparison of optical and
infrared outflow and accretion diagnostics (including profiles
of [O I] $\lambda$6300, H$\alpha$ and \heopt) for stars common to
both studies. The current sample thus includes 30 of the 31 CTTS
in HEG and BEK, identified in Table~\ref{t.sample} by an entry
for veiling at $\lambda$5700 \AA\ (r$_V$). Also included in
Table~\ref{t.sample} are disk accretion rates determined from
veiling, spanning ${\dot M}_{acc}$ $\sim 10^{-6}$ to $10^{-9}
M_{\odot} yr^{-1}$, mass outflow rates from forbidden lines,
spanning ${\dot M}_{w}$ $\sim 10^{-7}$ to $10^{-10} M_{\odot}
yr^{-1}$ and the corresponding ratio of ${\dot M}_w/{\dot M}_{acc}$.

We have determined both the 1~\micron\ veiling $r_Y$, defined as
the ratio of excess to photospheric flux in the $Y$ band, and
residual emission profiles at \helium\ and \pgamma\ for each of
our objects. Our method for veiling determination follows
\citet{hgn89}, matching an echelle order of an accreting star to
that of a non-accreting standard of comparable spectral type to
which has been added a constant continuum excess, i.e. the
veiling, that reproduces the depth of the photospheric features
in the accreting star. We used an interactive IDL routine kindly
provided by Russel White to accomplish the requisite velocity
shift of the standard via cross-correlation and the requisite
rotational broadening of the standard before determining the
required level of continuum excess. When a good match
between the accreting star and the veiled standard is found,
subtraction of the template yields a
zero-intensity residual spectrum for the accreting star except
in the vicinity of lines arising from accretion-related
activity. Integrated equivalent widths for spectral
lines were determined with
an IDL routine kindly provided by Catrina Hamilton.

\begin{figure}
\plotone{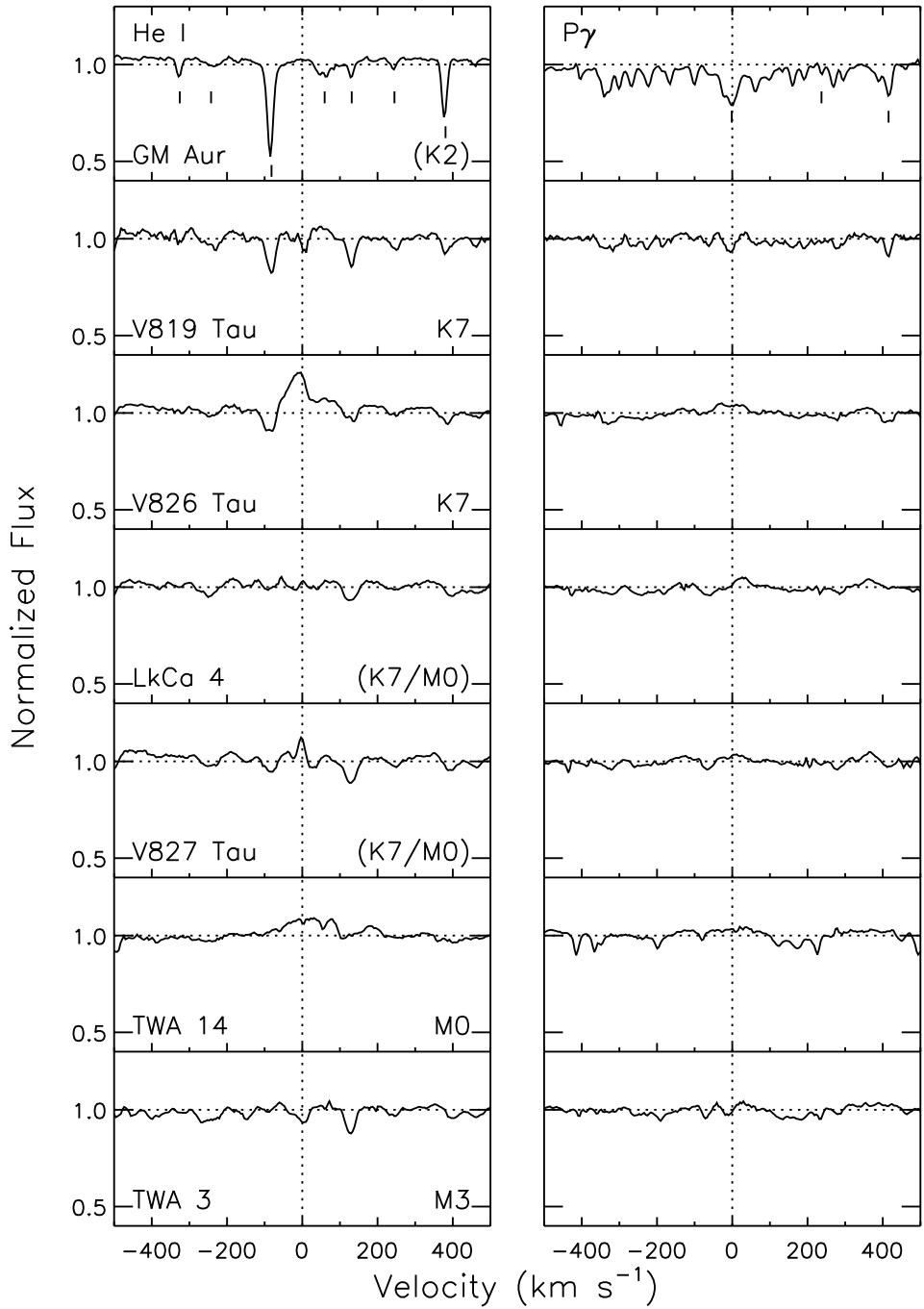}
\figcaption{Spectral regions of \helium\ and \pgamma\ for
6 WTTS plus GM Aur (see Table 1). Photospheric
features  marked with vertical lines are identified in the text,
except for the numerous features from CN 0-0 near
\pgamma. Ordering is by {\it apparent}
spectral type, given in parentheses when it differs from published
values. Fluxes are normalized to the continuum, velocities measured
relative to the stellar photosphere, and spectra are unbinned (i.e.
at the instrumental resolution.)
\label{f.wtts}}
\end{figure}

In Figure~\ref{f.wtts} continuum normalized spectra for each of
the 6 WTTS (LkCa 4, TWA 3, TWA 14, V819 Tau, V826 Tau, and V827
Tau) are shown along with the CTTS GM Aur. Spectra in $\pm$ 500
\kms\ windows are centered on the wavelength of the \helium\ and
\pgamma\ lines, at rest with respect to the star as determined
from the photospheric lines. In each region several photospheric
features are marked with vertical lines in the upper spectrum of
Figure~\ref{f.wtts} (GM Aur). In the \helium\ region the 3
strongest photospheric features are \ion{Si}{1} ~$\lambda
10830.1$, \ion{Na}{1}~$\lambda 10837.8$ and \ion{Si}{1}~$\lambda
10846.8$. In the \pgamma\ region the 3 marked lines correspond
to \ion{H}{1}~$\lambda 10941.1$ (\pgamma), \ion{Fe}{1} ~$\lambda
10949.8$ and \ion{Mg}{1} ~$\lambda 10956.3$. The remaining
numerous absorption features in that region are part of a broad
CN 0-0 band, not noise. V826 Tau is a spectroscopic binary
\citep{mundt} and in our data the photospheric lines are double
with a velocity separation of 21 \kms.

None of the 6 WTTS or GM Aur have any excess continuum emission
($r_Y < 0.05)$, and only two of them have weak emission at
\helium\ or \pgamma\ above an upper limit of $W_{\lambda} <$ 0.2
\AA. Both lines are weakly emitting in V826 Tau, with
$W_{\lambda}$ = 0.8 \AA\ for helium and $W_{\lambda}$ $\le 0.3$
\AA\ for \pgamma. TWA 14 has only helium in emission, with
$W_{\lambda}$ = 0.6 \AA. The emission equivalent widths for
these 2 stars are several times weaker than the weakest emission
found for the CTTS. Optical \heopt\ emission was also found to
be rare among WTTS, seen in only 3 out of 10 WTTS by BEK,
including V826 Tau. Both the optical and infrared helium
emission may arise in an active chromosphere or a very weak
accretion flow.

In earlier studies GM Aur has shown the hallmarks of an accreting
CTTS -- optical veiling, \ion{O}{1} emission and strong
H$\alpha$ (HEG, Gullbring et al.
1998). However, in our NIRSPEC data this is the only one of the
39 CTTS that shows no line emission in \helium\ or \pgamma\
($W_{\lambda} < 0.2$ \AA), which we interpret as a sign that
accretion has temporarily ceased. Its weak \pgamma\ absorption
feature is similar in strength to that from the photosphere of
an early K star \citep{hink}. The spectral energy
distribution for GM Aur has been interpreted as indicative of a
large (several AU) inner gap in its disk \citep{rice}, so the
absence of CTTS characteristics at 1 ~\micron\ in November 2002
might indicate a weakening or halt of accretion through the inner
gap. Because GM Aur had no detectable line emission or continuum excess, we
treated it as a WTTS and adopted it as a spectral template for
veiling. The spectral type for this star in the literature
ranges from K3 to K7. Is is certainly warmer than K7 in
our spectra, with photospheric features similar to stars
classified as K2-K3 (CW Tau and UX Tau). We also note that two
of other WTTS, TWA 3 (Hen 3-600A) and TWA 14, have been modeled
as possessing extremely small disk accretion rates of ${\dot
M}_{acc}$ $=5\times 10^{-11} M_{\odot} yr^{-1}$ on the basis of
their \halpha\ lines \citep{mcbh,muz01}. However, since the
derived accretion rates are so small and neither shows
significant line or continuum emission in our 1~\micron\ spectra
we have included them with the WTTS and adopted TWA 3 as a
veiling template.

The final palette of WTTS spectral templates for determination of
veiling and residual emission profiles in the CTTS included only
5 stars, after eliminating V826 Tau and TWA 14. Although 4 of
these are identified as K7/MO in the literature, the relative
strengths of their Ti, Cr and Si lines led us to sequence them
as shown in Figure~\ref{f.wtts}.  These correspond
to effective spectral types of K2, K7, K7, K7/MO and M3
for veiling determination. With this sequencing our modest number of
WTTS templates provided good veiling determinations for the CTTS
with the possible exception of the 2 solar type stars,
GW Ori and SU Aur. For these
stars we assigned SU Aur $r_Y = 0$, since its line depths
were comparable to our K2 template after the appropriate
rotational broadening, and from there we found that GW Ori, which has
a comparable $v\sin i$ to SU Aur, required $r_Y = 0.3$. In
addition there were 3 stars with a middle to late K spectral
type (CI Tau, HN Tau, and RW Aur A) which appeared to be
intermediate between our K2 or K7 templates so we took the
unorthodox step of creating a synthetic template from averaging
the spectra of these two. It provided an excellent match for the
relative line depths of these 3 stars, and we felt the veiling
derived with the synthetic template was reliable.

The $Y$-band veiling, $r_Y$, for the 38 CTTS with \helium\ and
\pgamma\ line emission is listed in Table~\ref{t.kin}. Stars are
ordered by veiling, which ranges from $r_Y = 0 - 2$. Veiling is
first determined independently in each of the two photospheric
orders, and then the average value is adopted and applied to the
order with the \helium\ and \pgamma\ lines. Subtraction of the
veiled standard from this order yields both residual emission
line profiles for these two lines and a third assessment of the
veiling. We found an excellent agreement for veilings determined
from the 3 orders. The error for each veiling measurement is the
largest deviation from the adopted value that yields a fit
indistinguishable from the best fit, which is typically larger
than the dispersion among the orders. When $r_Y$ is less than or
equal to 0.5, the veiling error is approximately $\pm 0.05$. For
larger veilings, the error is approximately $\pm 0.1$.  We note
that the veiling measurements reported here for the 6 stars that
also appear in Edwards et al. (2003) supercede the earlier estimates,
which were not evaluated with the same degree of thoroughness.

\begin{figure}
\plotone{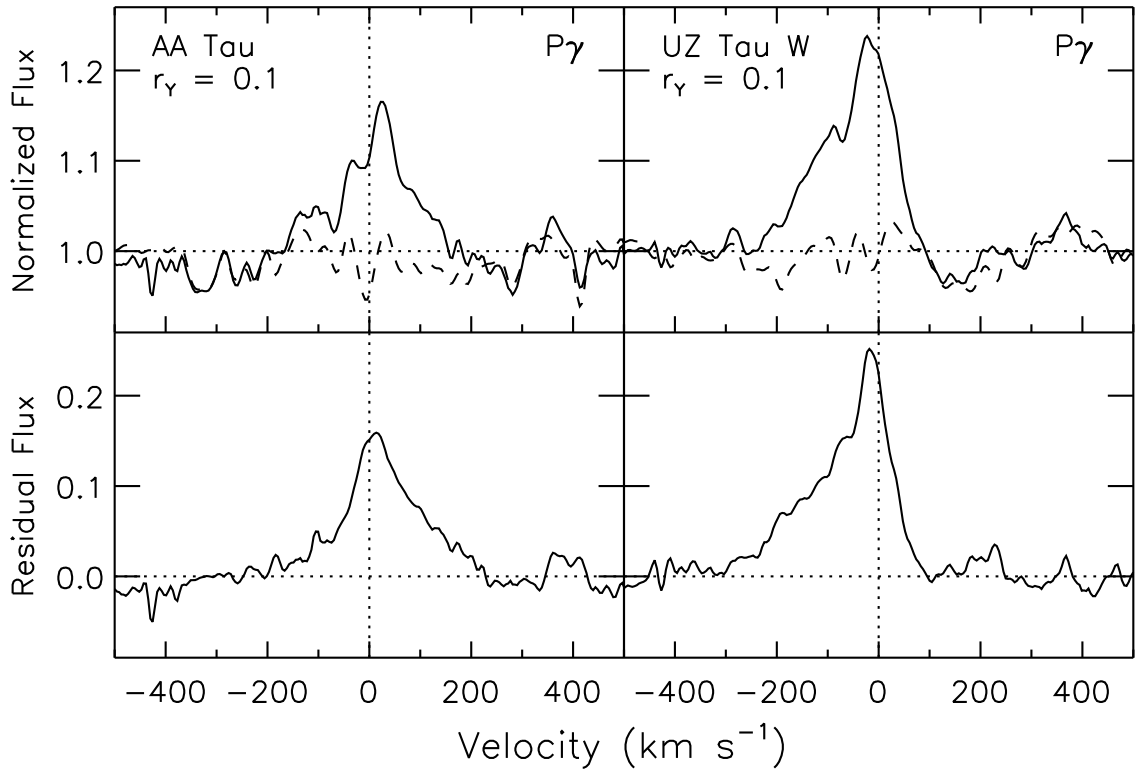}
\figcaption{Demonstration of {\it residual profile} determination at
\pgamma\ in AA Tau and UZ Tau W.  The top row shows fluxes relative
to the continuum for the CTTS (solid line) and for the veiled template
(dashed line). The resultant residual profiles after subtraction of the
template from the star are shown in the lower row. Templates are V819 Tau
(K7) for AA Tau and TWA 3(M3) for UZ Tau W. Velocities are relative to the
stellar photosphere.
\label{f.residual}}
\end{figure}

The process of creating residual profiles results in an improved
definition of the emission lines in stars with low veiling,
particularly in the line wings. We illustrate this in
Figure~\ref{f.residual} at \pgamma\ for two stars with $r_Y =
0.1 $, AA Tau and UZ Tau W. Note the effect on the line shape,
especially how the wings in the residual profiles are more
clearly defined when the photospheric features have been
removed.

The residual profiles for \pgamma\ and \helium\ for 38 CTTS
 are presented in Figure~\ref{f.pg_ctts} and
Figure~\ref{f.he_ctts}, respectively. The 39th CTTS, GM
Aur, is excluded since it was not detected in either line.
Nine CTTS were observed
more than once, and we show the multiple profiles for \pgamma\
and \helium\ in Figure~\ref{f.pg_repeat} and
Figure~\ref{f.he_repeat}, respectively. Most multiple
observations are separated by days, except for DG Tau and HL
Tau, which are separated by a year. Table~\ref{t.kin} presents
measurements of the \pgamma\ and \helium\ features in the CTTS
including kinematic information and emission and absorption
equivalent widths. Among the 9 stars with more than one
observation, the profiles and their kinematic parameters do not
change appreciably with the exception of AA Tau. For these 9
stars, the spectrum included in the {\it reference sample} is
identified with an asterisk, representing one spectrum for each
star, and is used for all statistical analyses in figures and
text.

In both the figures and Table~\ref{t.kin} the spectra are
arranged in order of decreasing 1~\micron\ veiling. The table
separates the stars into groups corresponding to 3 levels of
veiling:  high  ($r_Y \geq 0.5$; 9 stars),
medium  ($0.3 \le r_Y < 0.5$; 11 stars) and low
($r_Y \le 0.2$, 18 stars). The low veiling stars are further
divided into 2 subgroups of 9 stars each, according to the width
of the \pgamma\ feature. Average values for kinematic properties of the
profiles in each group are identified. Discussion of the profiles and their
kinematic properties follows in subsequent sections.

\section{ACCRETION DIAGNOSTICS: \pgamma\ AND 1~\micron\ VEILING}

Our NIRSPEC data allows us to evaluate two quantities expected to
be sensitive to the disk accretion rate: \pgamma\ equivalent width and the
1~\micron\ veiling, $r_Y$. The \pgamma\ line emission is predicted to
arise primarily in magnetospheric accretion columns \citep{mch98,mch01}
and the continuum
excess from accretion luminosity, e.g. from accretion shocks at the
base of magnetic funnel flows \citep{cg98}. These diagnostics can be used to
test magnetospheric accretion models and to
establish the extent to which our prime wind diagnostic, \helium, is
influenced by accretion. We detect \pgamma\ emission in 38 of
the 39 CTTS and 1~\micron\ veiling in 31 of the 39 CTTS. The
sole \pgamma\ non-detection is GM Aur, which as discussed in the
previous section appears to be in a quiescent non-accreting
state, and we treat it as a WTTS for the remainder of this
paper (see Figure~\ref{f.wtts}). In this section we focus on the properties of the
\pgamma\ emission and its relation to veiling. We will discuss
the implications of the profile morphology for magnetospheric accretion
models in Section 6.1 and examine the spectral energy distribution of
the veiling and its relation to disk accretion rate in Section 7.
For the intervening sections we simply use
$r_Y$ as a quantity likely to be proportional to the
instantaneous disk accretion rate.

\begin{figure}
\plotone{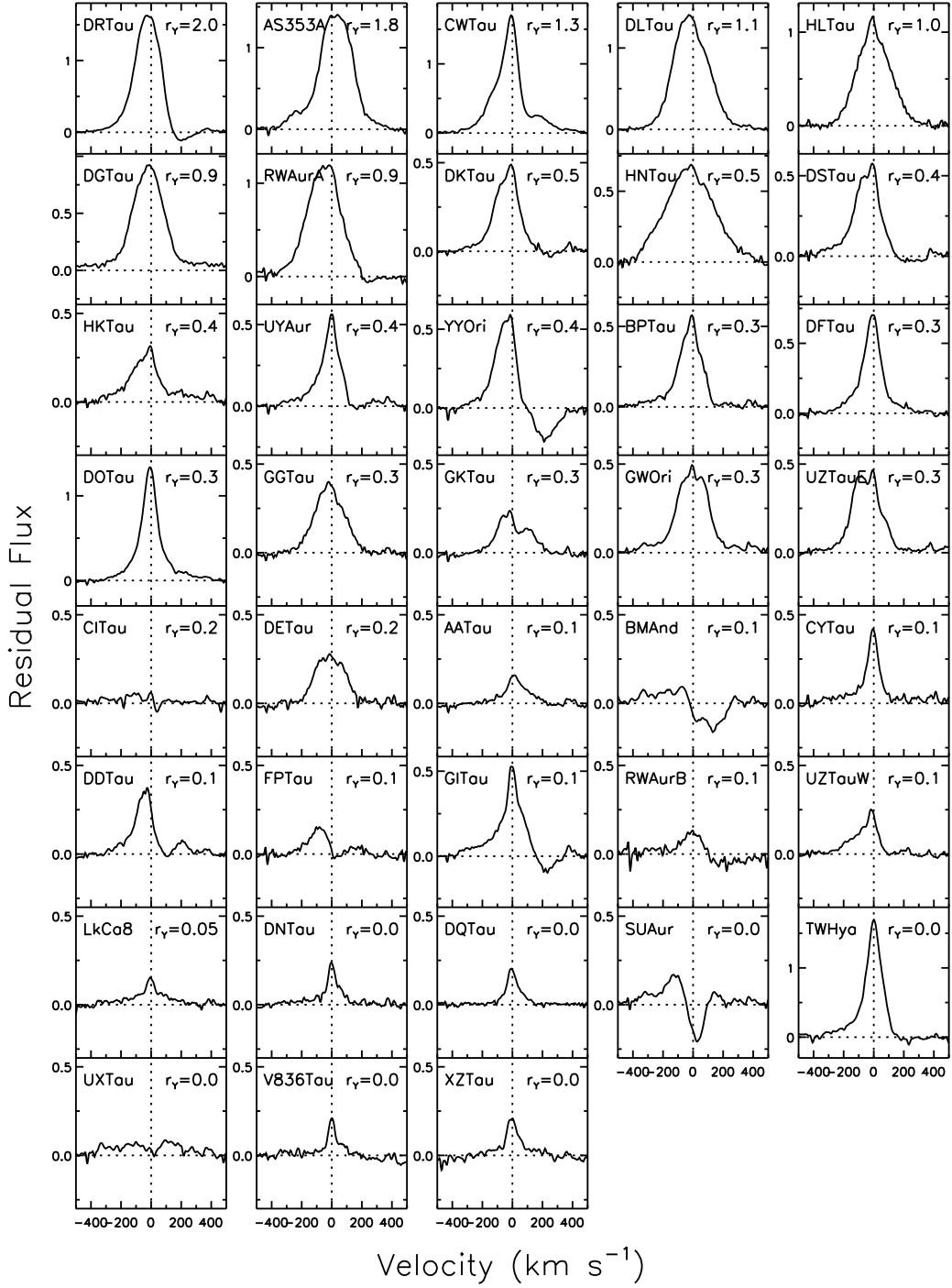}
\figcaption{{\it Residual} \pgamma\ profiles from the
reference sample for 38 CTTS
(listed in Table 1), ordered by decreasing 1~\micron\ veiling $r_Y$.
Velocities are relative to the stellar photosphere and spectra are
plotted with 3 pixel binning.
\label{f.pg_ctts}}
\end{figure}

The \pgamma\ profiles for 38 CTTS are predominantly in emission;
none show blueshifted absorption from a
wind while 24\% (8 from the reference
sample in Figure~\ref{f.pg_ctts}: BM And, DR Tau, DS Tau, GI Tau,
RW Aur, RW Aur B, SU Aur, YY Ori plus AA Tau in
Figure~\ref{f.pg_repeat}) show redshifted absorption below the
continuum, suggesting infall of accreting material in a
magnetospheric funnel flow. The \pgamma\ emission and absorption
equivalent widths are listed for each observation in
Table~\ref{t.kin}. The emission equivalent widths range from a
high of 18 \AA\ (AS353 A) to a low of 0.6 \AA\ (CI Tau, V836
Tau) and the absorption equivalent widths from a high of 1.1
\AA\ (YY Ori) to a low of 0.2 \AA\ (DS Tau and RW Aur A).

We find a remarkably good correspondence between the \pgamma\
emission equivalent width and $r_Y$, shown in the top panel of
Figure~\ref{f.eqwact}. Following standard practise, in the
figure the measured equivalent width $W_{\lambda}$ is multiplied
by the factor $(1 + r_Y)$, which renormalizes it to the
photosphere, rather than photosphere plus veiling continuum
\citep{bb90, BEK}. The figure shows a tight correlation between
\pgamma\ and $r_Y$ for the 31 CTTS with measurable veiling, with
a linear correlation coefficient of 0.97 between the logarithm
of the veiling and the logarithm of the veiling-corrected
\pgamma\ emission. The figure exaggerates the separation in
veiling between the 11 CTTS with the lowest measured veiling,
$r_Y \sim$0.1, and the 7 CTTS non-detections, $r_Y <$ 0.05,
which were not included in the determination of the correlation
coefficient. However all but one of the non detections have
\pgamma\ strengths comparable to those with the lowest veilings
so they effectively exhibit the same correlation between
\pgamma\ emission and veiling. The sole exception to the
correlation is TW Hya, which has no detected veiling but robust
\pgamma\ emission comparable to CTTS with $r_Y \sim 0.3$. The
tight relation between $W_{\lambda}$ (\pgamma) and $r_Y$ is
reminiscent of, although better defined than, others reported
between permitted emission lines and either veiling or disk
accretion rates determined from veiling (\cite{bb90} for
\halpha; \cite{mhc98b} for \halpha\ and P$_{11}$; \cite{mhc98c}
for Br$\gamma$ and P$\beta$; \cite{dop03} for Br$\gamma$).

Similarly, the morphology of the \pgamma\ profiles
shown in Figure~\ref{f.pg_ctts} is reminiscent of P$\beta$ and
 Br$\gamma$ profiles \citep{folha01,mhc98c}, with broad
single peaks sometimes accompanied by redshifted absorption
below the continuum. The profiles cover a range of
line widths and centroids, which we characterize by
measuring 3 kinematic properties: (1) emission
centroids which range from \vc\ = -150 to +29 \kms
(2) maximum blue wing velocities which range from
$V_{bwing}$ = -400 to -100~\kms; and (3) line widths measured from
the stellar rest velocity to the blueward half-intensity point
which range from \vbhi\ = -250 to -26 \kms. The latter two
focus on emission blueward of the stellar
rest velocity in order to avoid complications from redshifted
absorption:. Measurements of $V_{bwing}$, \vbhi\, and \vc\
are plotted against veiling in Figure~\ref{f.pgkin} for all but two stars
in the reference sample and listed in Table~\ref{t.kin} for all observations.
The 2 stars omitted in the figure (CI Tau, UX Tau) have such weak and amorphous \pgamma\
emission that  kinematic features could not be reliably established.

The kinematic data for \pgamma\ suggest a relation that has not
previously been reported: a connection between \pgamma\ profile
morphology and veiling. We illustrate it more directly in
Figure~\ref{f.pgcat} where we superpose \pgamma\ profiles in
groups segregated by veiling. The top panel displays profiles
for 9 CTTS with high veiling ($r_Y \geq 0.5$), the second panel
for 11 CTTS with medium veiling ($0.3 \le r_Y < 0.5$), and last
two panels for 18 CTTS with low or absent veiling ($r_Y \le
0.2$). The low veiling group is divided into 2 subcategories
according to the width of their \pgamma\ profiles: 9 CTTS in the
3rd panel have narrow \pgamma\ lines with \vbhi\ $<$ 60 \kms and
9 CTTS in the 4th panel have \pgamma\ lines in excess of this
width.

The trend in profile morphology with veiling is exhibited by the
29 stars in the top 3 panels (i.e. excluding the 9 low veiling
stars with broad \pgamma\ profiles, which will be discussed
below). For these 29 stars, \pgamma\ profiles are normalized to
their peak intensities (given for each star in
Table~\ref{t.kin}) to facilitate comparison of profile
morphology independent of line strength. Among these 29 stars,
the mean linewidth blueward of the stellar velocity \vbhi\
decreases from 135 \kms\ to 40 \kms\ among the 3 veiling groups,
while the mean centroid velocity decreases from \vc\ $\sim$20 \kms\
to $\sim$0 \kms. Remarkably, those in the low veiling subgroup with narrow
\pgamma\ (\vbhi\ $\le 60$ \kms ) all have rather similar
profiles after normalization, showing central peaks with
an average \vbhi\ = $-40 \pm 10$ \kms, an average \vc\ = $2
\pm 8$ \kms. Most also have a two-component structure suggesting a strong narrow
core and a weak broad base. When we compute an average \pgamma\ profile for
these 9 low veiling stars it is well fit with two gaussians, one a narrow
component centered at -2 \kms\ with a FWHM of 52 \kms\ and the
other a broad component centered at +6 \kms\ with a FWHM of 176 \kms.

Although the trend in \pgamma\ profile morphology to decrease in
width and centroid velocity with decreasing veiling
is suggestive it does not include 9 stars with low veiling (24\% of the sample).
The separation of the low veiling stars into two groups based on
\pgamma\ line width was effected in order to illustrate the
remarkable uniformity for half of the low veiling profiles. The
other 9 low veiling stars with broad \pgamma, superposed in the
bottom panel of Figure~\ref{f.pgcat}, show an unusual diversity
in their morphology. Three (BM And, CI Tau, UX Tau) have broad
amorphous emission and low peak intensity ($<$10\% of the continuum flux). Two
(BM And and SU Aur) have exceptionally
deep and broad redshifted subcontinuum absorption penetrating to 50\% of the
1~\micron\ continuum with emission peaking considerably blueward
of the other stars (\vc\ $<$ -120 \kms). Along with these 4
stars which are completely unlike the rest of the sample, the
additional 5 stars in this subgroup (DE Tau, DD Tau, FP Tau, RW
Aur B, UZ Tau W) are more typical but are included here because
their breadths and blueshifts are larger than those with uniform
narrow \pgamma\ profiles. (We note that unlike the other
superposed profiles in Figure~\ref{f.pgcat}, the 9 stars in this
subgroup are plotted in residual intensity units, reflecting
their true height above the stellar continuum rather than
normalized to peak intensity, to avoid mis-representing the
amorphous profiles with very low peak intensities.)

In summary, our key findings for \pgamma\ and veiling are (1) a
robust correlation between emission equivalent width and
veiling; (2) a remarkably consistent profile morphology among
half of the stars with low veiling showing a two component
structure with narrow centered cores and a weak broad base; and
(3) a progression in profile morphology with veiling for a
sizeable fraction, but not all, of the sample. Systematic
variations with veiling of line width and centroid velocity have
not, to our knowledge, been previously reported for hydrogen
lines in CTTS. We discuss the implication of the veiling
dependence of \pgamma\ profiles for magnetospheric formation in
Section 6.1.

\section{WIND DIAGNOSTIC: \helium}

Our NIRSPEC survey probes the innermost wind region in accreting
stars with the \helium\ line. The uniqueness of this probe
derives from the metastability of its lower level
($2s^3S$), which, although energetically far above (20
eV) the singlet ground state, is radiatively isolated from it.
Whether this metastable level is populated by recombination and
cascade or by collisional excitation from the ground state, it
will become significantly populated relative to other excited
levels owing to its weak de-excitation rate via collisions to
singlet states, making it an ideal candidate to form an
absorption line. This absorption is essentially a
resonant-scattering process since the $\lambda$10830 transition
($2p ^3P^o - 2s ^3S$)
is the only permitted radiative transition from its upper state
to a lower state and the electron density is unlikely to be so
high as to cause collisional excitation or de-excitation. This
sets up an ideal situation to probe outflowing gas in
absorption, provided the conditions are right to excite helium
20 ev above the ground state.

\begin{figure}
\plotone{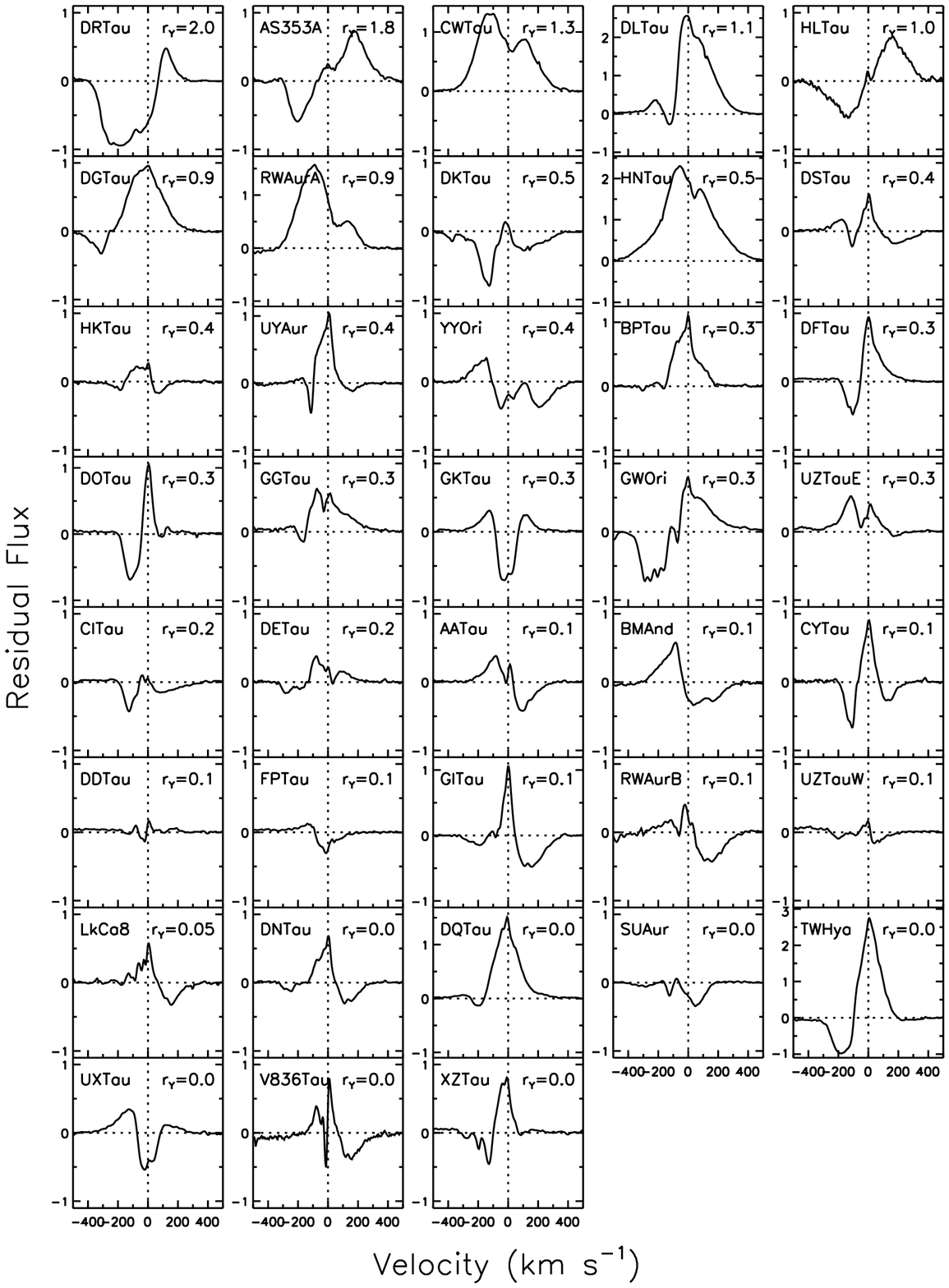}
\figcaption{{\it Residual} \helium\ profiles from the reference sample
for 38 CTTS ordered by decreasing 1~\micron\ veiling $r_Y$.
Velocities are relative to the stellar photosphere and spectra are
plotted with 3 pixel binning.
\label{f.he_ctts}}
\end{figure}

\begin{figure}
\plotone{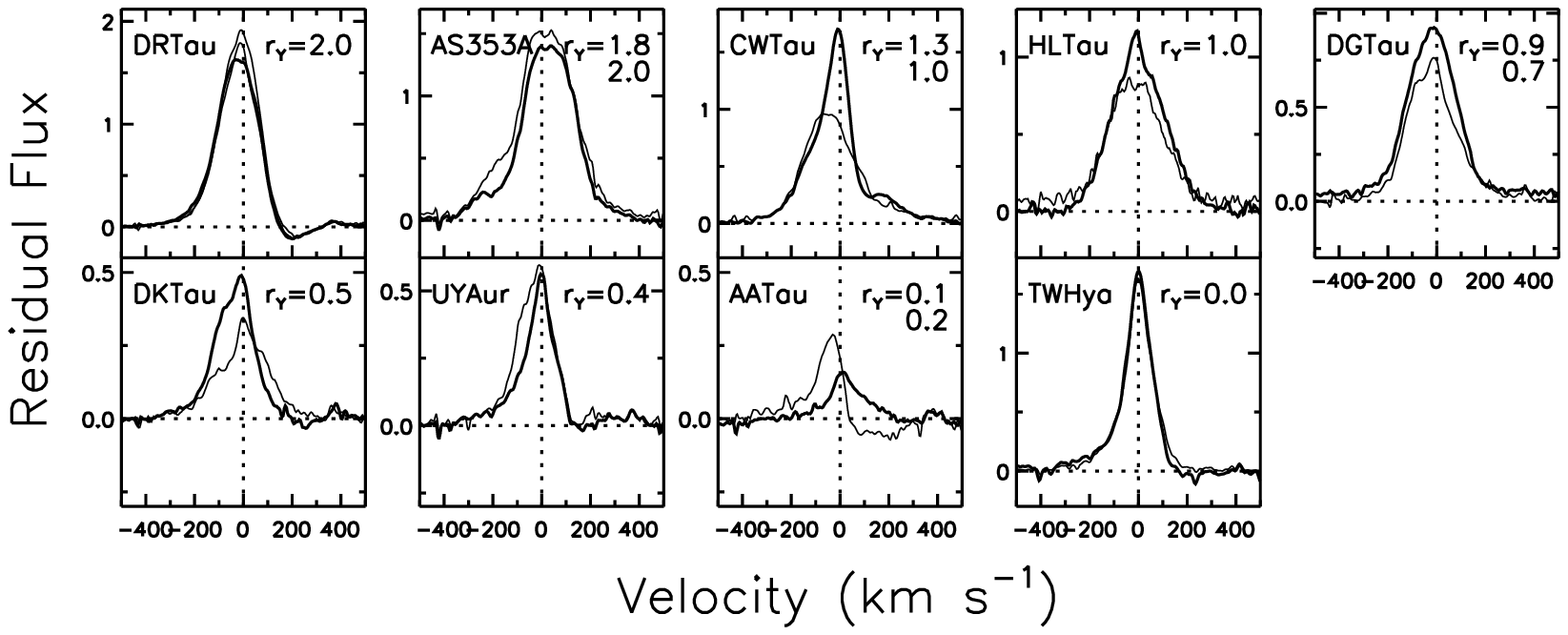}
\figcaption{Multiple observations of \pgamma\ residual profiles for 9 CTTS.
For each object, the spectrum from the reference sample is denoted by a heavier line.
\label{f.pg_repeat}}
\end{figure}

\begin{figure}
\plotone{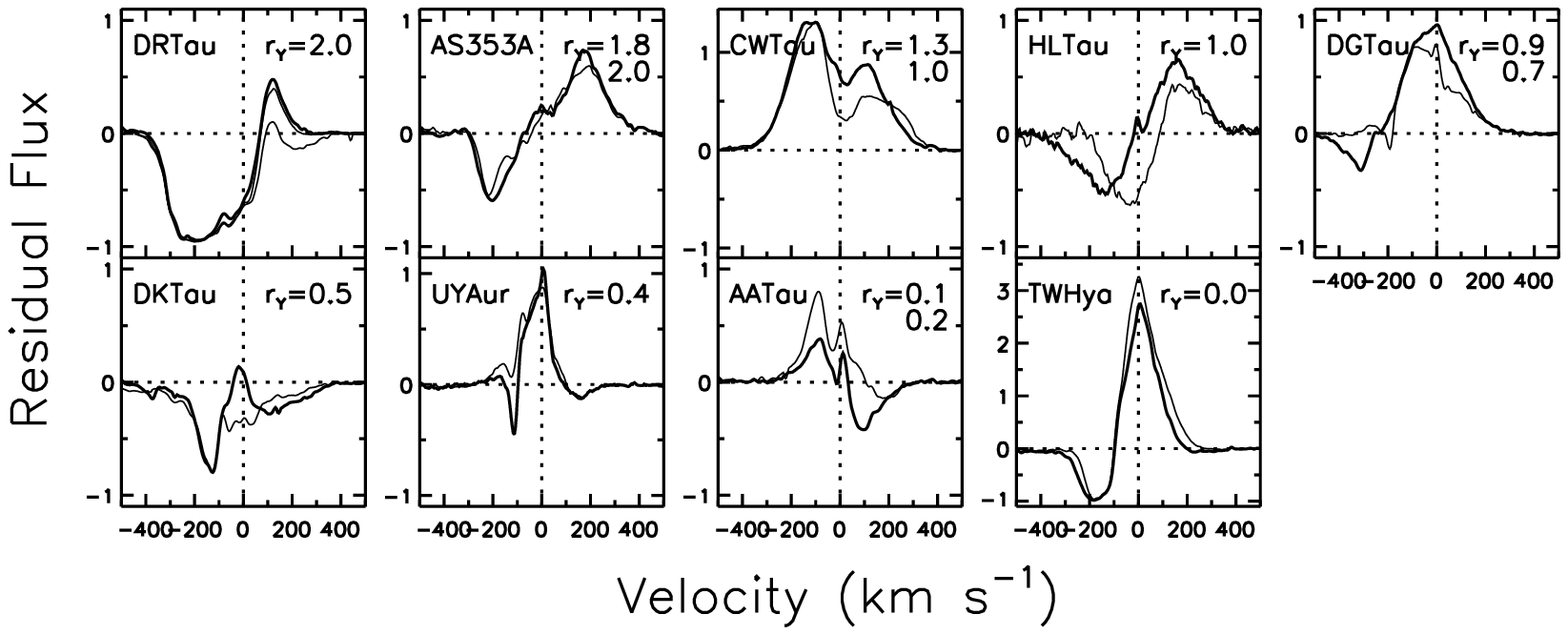}
\figcaption{Multiple observations of \helium\ residual profiles for 9 CTTS.
For each object, the spectrum from the reference sample is denoted by a heavier line.
\label{f.he_repeat}}
\end{figure}

\begin{figure}
\plotone{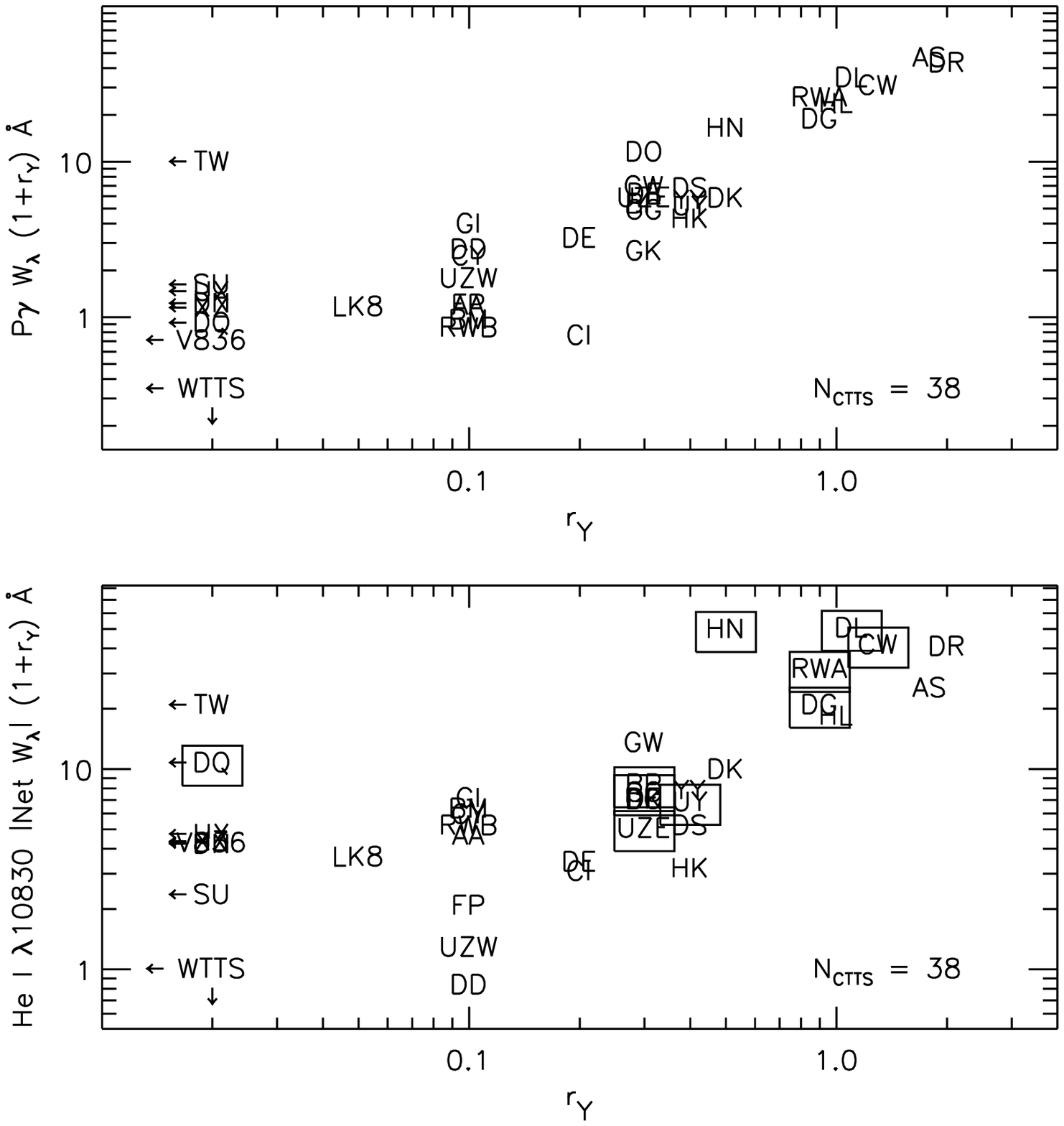}
\figcaption{\emph{Top panel}: Veiling-corrected \pgamma\ emission equivalent width versus
$Y$-band veiling for 38 CTTS from the reference sample. \emph{Bottom panel}:
Veiling corrected \helium\ activity index (sum of absolute values of emission plus
absorption equivalent widths) versus $Y$-band veiling for the same stars. In the
\helium\ panel a box around the name of the star indicates its equivalent width is
primarily in emission (see text). Veiling non-detections are shown at $r_Y=0.02$ for clarity.
The location of the 6 WTTS plus GM Aur are also indicated.
\label{f.eqwact}}
\end{figure}

\begin{figure}
\plotone{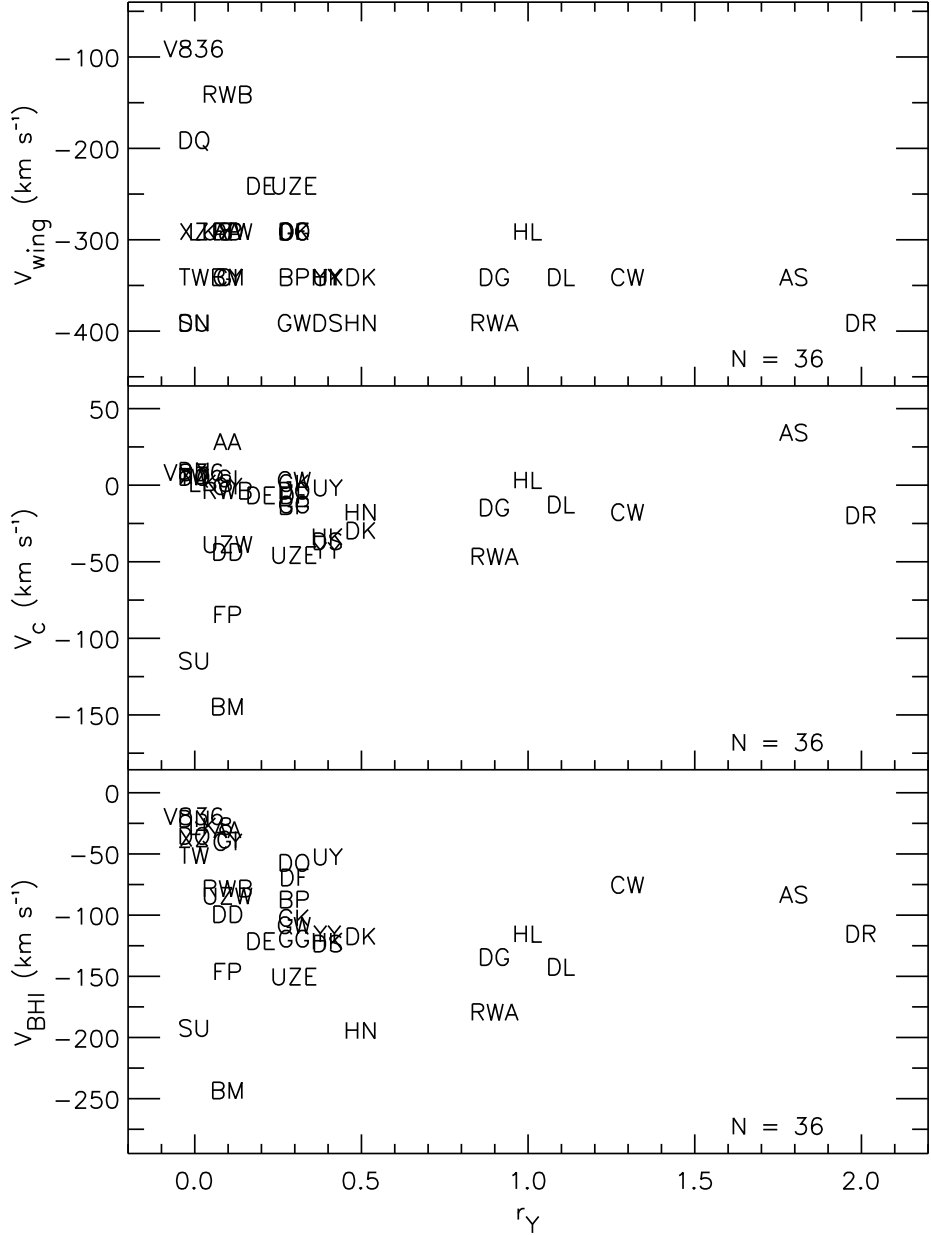}
\figcaption{Kinematic parameters $V_{bwing}$, \vbhi, and \vc\ for \pgamma\ profiles
plotted against veiling.
\label{f.pgkin}}
\end{figure}

\begin{figure}
\plotone{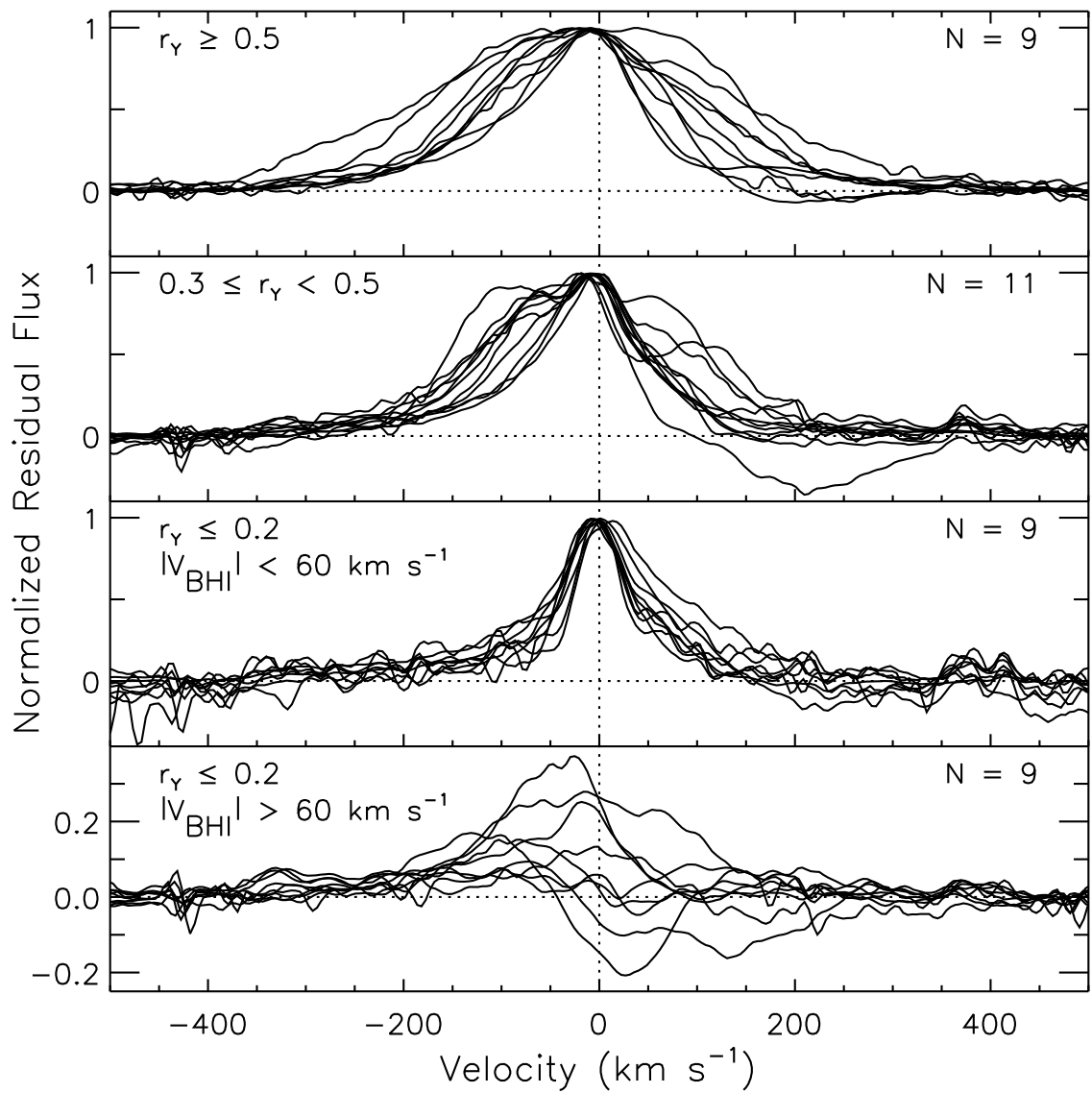}
\figcaption{Superposed \pgamma\ residual profiles from the reference sample
grouped by veiling, with the lowest veiling group subdivided into narrow and wide
profiles (panels 3 and 4). In the top 3 panels profiles are normalized to their
peak intensities. In the lower panel the residual profiles are not normalized in order
not to distort the very weak amorphous emission characterizing 3 stars.
\label{f.pgcat}}
\end{figure}

We find \helium\ features in 38 of the 39 CTTS in our sample; the
exception is GM Aur (see Figure~\ref{f.wtts}), suspected to be in a quiescent
non-accreting state. Line profiles for each star, ordered by
decreasing $r_Y$, are shown in Figure~\ref{f.he_ctts} and additional
observations of the 9 stars with multiple observations are in
Figure~\ref{f.he_repeat}. Equivalent widths and kinematic properties
are listed in Table~\ref{t.kin}, sorted by 4 groups corresponding
to high, medium and low veiling, with the 18 low veiling objects
additionally split two kinematic subgroups based on \pgamma\ line widths,
\vbhi.

The predicted sensitivity of \helium\ to absorption is born out
by our observations, where absorption
below the 1~\micron\ continuum is seen in 34 of the 38 stars (89\%).
Although blueshifted subcontinuum absorption from outflowing gas
is most common (27 of 38 stars, 71\%) redshifted subcontinuum absorption
from infalling gas is also frequent (18 out of 38 stars, 47\%). In 2
stars (5\%; GK Tau, UX Tau), the subcontinuum absorption is centered
on the stellar rest velocity and is flanked by emission on both
sides. Our identification of the presence of blue, red, or
centered subcontinuum absorption in each profile is listed in
Column 14 of Table~\ref{t.kin} by the letters {\it b, r, c}. Of
the 4 stars with no subcontinuum \helium\ absorption (CW Tau, RW
Aur A, HN Tau, BP Tau) the profiles have asymmetries suggesting
either redward or blueward absorption that does not penetrate
the 1~\micron\ continuum.

Because subcontinuum absorption is often a sizeable contributor
to the \helium\ line, a comparison between emission equivalent
width and veiling, which yielded a well defined relationship for
\pgamma, is not meaningful for \helium. While the emission
equivalent width above the continuum ranges from $W_{\lambda}$ =
0 to 30.1~\AA, the absorption equivalent width below the
continuum in turn ranges from $W_{\lambda}$ = 0 to 10.7~\AA\ so
that a strong helium feature may have a large positive or
negative equivalent width, or may have a net equivalent width
close to zero. To examine the dependence of the strength of
\helium\ with veiling, we define a {\it helium activity
index} as the sum of the absolute value of the equivalent width of the
emission above the continuum ($E$) and the absorption below the continuum
($A$). This activity index ($E+A$) ranges from $\sim 1-30$ and
is plotted against $r_Y$ in the lower panel of Figure~\ref{f.eqwact}.
It shows a trend of increasing with veiling,
corresponding roughly to that seen for \pgamma, although with
considerably more scatter. We have separately indicated the
helium profiles that are primarily in emission ($E-A/E+A > 0.5$)
since the activity index for these stars is dominated by the
emission equivalent width, which is directly analogous with
the relation found for \pgamma.
The observed trend relating the helium activity index
and veiling indicates that
helium excitation is correlated with disk accretion rate.
 
\begin{figure}
\plotone{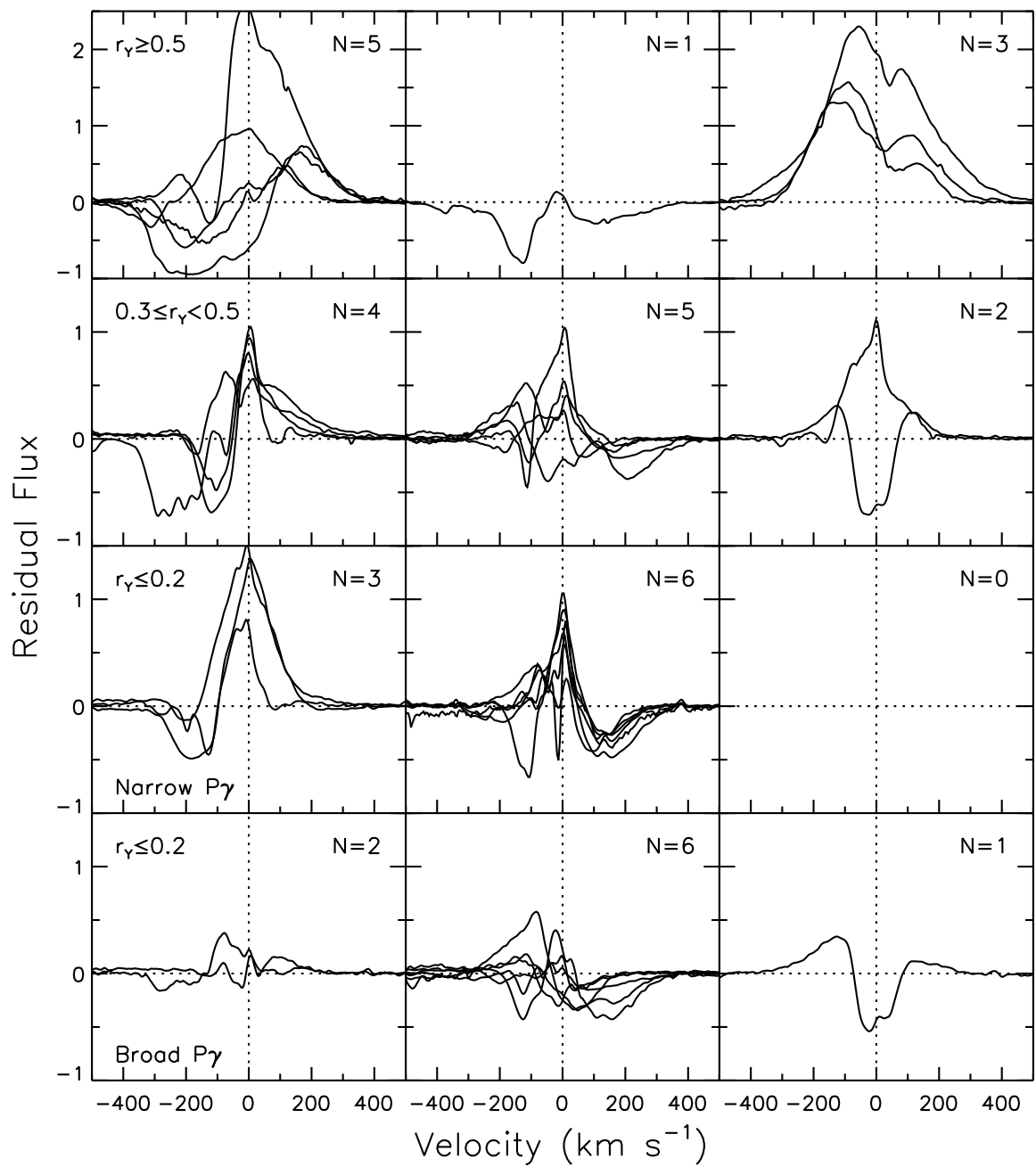}
\figcaption{Superposed \helium\ residual profiles grouped by
1~\micron\ veiling (horizontal rows) and general morphology (vertical
columns).  Veiling groups are identical to figure~\ref{f.pgcat}, with the
low veiling group subdivided by the width of \pgamma. Morphology groups are:
{\it left:} ``P-Cygni like'', {\it middle:} red absorption below the continuum
and {\it right:} other.
\label{f.hecat}}
\end{figure}

In order to appreciate the full diversity of helium profiles, we have
arranged them into a scheme based on a combination of both
veiling level and profile structure. This categorization is
shown in Figure~\ref{f.hecat}, where superposed \helium\
profiles are first grouped by the same 4 categories used for
comparing \pgamma\ profiles in Figure~\ref{f.pgcat} based on
veiling and the \pgamma\ line width (4 horizontal rows), and
then further sorted into 3 morphological classes (3 vertical
columns). All helium profiles are in the continuum-normalized
units of Figure~\ref{f.he_ctts}, with the exception of TW Hya
which was rescaled to half its peak intensity in order to facilitate
comparison with the other profiles in its category.

The 3 morphological groups portrayed in Figure~\ref{f.hecat}
are as follows: The left column contains 14 stars with profiles that fall
into the general category of ``P Cygni-like'', with subcontinuum
absorption that is blueward of all or most of the emission and with
 no redshifted absorption below the
continuum.  The middle column contains 18 stars with profiles
that do have redshifted absorption below the continuum, most of
which (13 stars) {\it also} show subcontinuum blueshifted absorption.
The profiles in the right column are from the 6 objects with
either central absorption (2 stars), or no subcontinuum
absorption (4 stars).

The ensemble of \helium\ profiles in Figure~\ref{f.hecat}
illustrates that each of the 3 general profile types (blueshifted
subcontinuum absorption, both blueshifted and
redshifted subcontinuum absorption, and central
or no subcontinuum absorption) can be found at all veiling
levels. However there is a tendency for broad P Cygni-like
profiles to be more common among stars of high veiling (5 out of 9
stars) while narrow emission coupled with
both blueshifted and redshifted absorption is more common among
stars with low veiling (12 out of 18 stars). Although our time
coverage is limited, among the 9 stars with more than one
observation we do not see any stars shifting among these profile
categories (see Figure~\ref{f.he_repeat}), suggesting that the profile
categories we observe are not primarily due to time variable behavior
among the CTTS but are intrinsic to each star.

Examination of the superposed profiles in Figure~\ref{f.hecat}
also shows that the kinematic structure of the subcontinuum blueshifted absorption
component is more diverse than that of the red absorption. We discuss the
implications of this diversity for wind formation in Section 6.2, and present
here a table with basic properties of the blueshifted absorption in
Table~\ref{t.bluabs}. This includes the equivalent width of the blue
absorption below the continuum, the continuum penetration depth,
the full velocity extent of the blue absorption and the corresponding
minimum and maximum velocity relative to the stellar rest velocity.
At one extreme is an object like DR Tau, where the \helium\
line is almost entirely in absorption, with a breadth of nearly
500 \kms, extending from -400 to +70 \kms\ relative to the star
and absorbing 95\% of the 1~\micron\
continuum over most of that velocity interval. At another extreme, an object
like V836 Tau has helium primarily in emission and the
narrow blueshifted absorption extends from only -30 to 0
\kms\ relative to the star, cutting through the emission feature
and penetrating the continuum to a depth of 30\%.

In summary, the prevalence in accreting T Tauri stars of
blueshifted He I $\lambda 10830$ absorption below the continuum
demonstrates the ubiquity of accretion powered winds arising in
the immediate vicinity of the star. However the diversity of
profile morphologies suggests that the nature of the winds
are not identical in all of these accreting stars, a topic we
return to in Section 6.2. We also find that \helium\ is
sensitive to infalling gas in magnetospheric accretion flows,
discussed further in Section 6.1.

\section{CLASS I  SOURCES}

In addition to the survey of 39 CTTS and 6 WTTS whose accretion
properties are well documented from previous studies, we
acquired NIRSPEC data for 3 young stellar objects, identified in
Table~\ref{t.class1sam}, whose spectral energy distributions
place them as Class I or borderline Class I/II objects. Two of
them, SVS 13 and IRAS 04303+2240, have disk accretion rates
comparable to those of the most active Class II CTTS, ${\dot
M}_{acc}$ $\sim 10^{-6} M_{\odot} yr^{-1}$ and are the driving
sources for HH Objects/jets and molecular outflows (references
given in Table~\ref{t.class1sam}). The third, IRAS 04248+2612,
is considerably less active with disk accretion and wind mass
loss rates several orders of magnitude smaller, comparable to a
typical CTTS \citep{wh04}.

The faintness of these embedded sources resulted in lower S/N
spectra than for the CTTS, where integration times of 300
seconds for IRAS 04303+2240 and 1200 seconds for SVS 13 and IRAS
04248+2612 yielded S/N of 60, 60, and 20, respectively). For
IRAS 04248+2612, the M5.5 photospheric features
showed  no evidence of veiling. The other two
objects have strong emission lines at the location of all the
photospheric features, preventing a direct measure of the
veiling. For them we estimated the veiling based on the relation
between \pgamma\ and $r_Y$ that we found for the CTTS. If they
follow the same relation, their \pgamma\ equivalent widths
indicate $r_Y \sim 1.8$ for IRAS 04303+2240 and $r_Y \sim
0.3$ for SVS 13.

\begin{figure}
\plotone{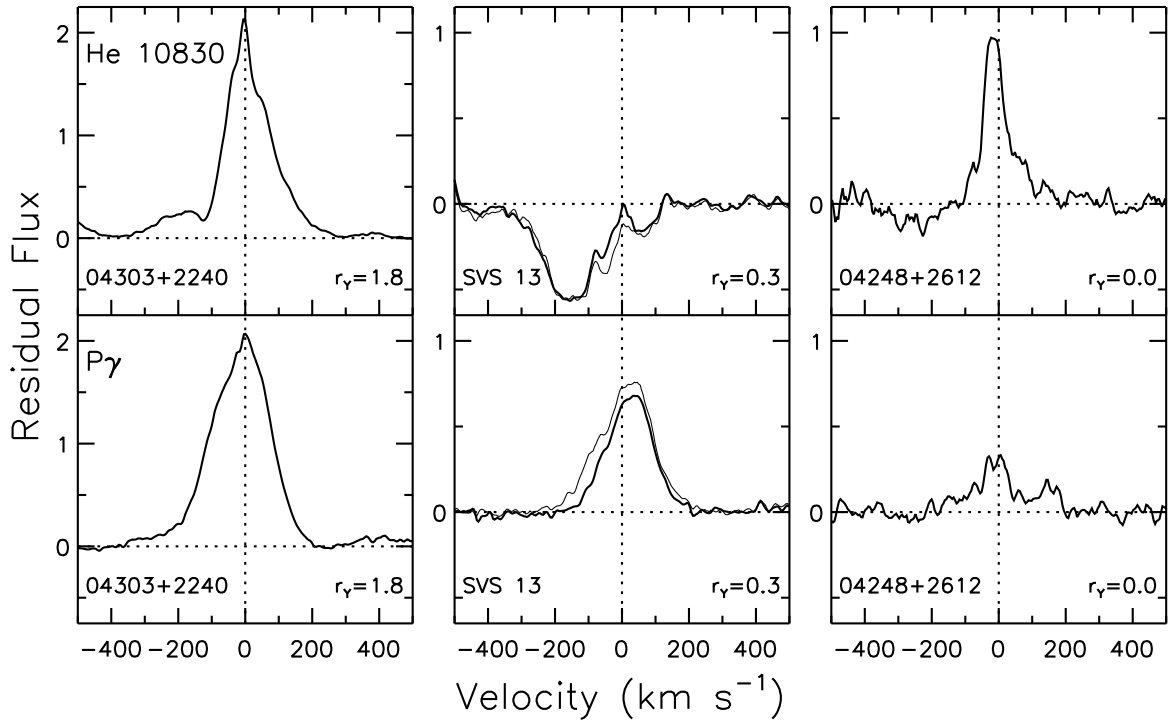}
\figcaption{Residual \helium\ and \pgamma\ profiles for the Class
I sources in order of decreasing 1~\micron\ veiling $r_Y$.
The 2 spectra for SVS 13 were taken a day apart; the reference spectrum ($r_Y=0.3$)
is shown with a heavier line than the other ($r_Y=0.4$). No photospheric features
were seen for SVS 13 or 04303+2240 and these profiles are referenced relative to the
velocity of ambient molecular material.
\label{f.class12}}
\end{figure}

The residual \helium\ and \pgamma\ profiles for these 3 stars are shown in
Figure~\ref{f.class12} and their line properties are given in
Table~\ref{t.class1kin}. Overall the profile morphologies for both
\pgamma\ and \helium\ are similar in character to those found among the CTTS,
with the exception of the \pgamma\ profile of the low accretion rate object
IRAS 04248+2612 which has a double rather
than single peaked structure. All 3 \helium\ lines show
evidence of winds in the form of blueshifted absorption, penetrating
the continuum in SVS 13 and IRAS 04248+2612 but not in IRAS 04303+2240.
The \helium\ profile for SVS 13 is very similar to that of the high
accretion rate CTTS DR Tau, in that both are almost entirely in absorption,
which is very broad and very deep, extending from the stellar rest velocity
to almost 400 \kms.

\section{DISCUSSION OF LINE PROFILES}

Our survey  of \helium\ and \pgamma\ lines in CTTS and 3 Class I sources
presents a rich new array of kinematic diagnostics of the star-disk interface
region in accreting stars. In this section we discuss how the profile
morphology for each of these lines gives insight into magnetospheric accretion
flows and accretion powered winds.

\subsection{Magnetospheric Contributions to \pgamma\ and \helium}

The presence of magnetospheric accretion flows in classical T
Tauri stars is inferred from general similarities between
observed and model profiles, particularly for hydrogen lines of
modest opacity \citep{mch01,hhc94,e94}. In the radiative transfer
models, line emission arises along the full length of an
accretion column characterized by a fairly uniform temperature,
connecting the truncated disk to the stellar photosphere. Hydrogen profile
characteristics reproduced in the models include broad single
peaked emission with a blue asymmetry and the appearance of
redshifted inverse P Cygni absorption (IPC) when the optical
depth is not too high and the inclination is favorable (i.e.
when the line of sight passes through both the infalling gas and
the hot accretion shock). In our NIRSPEC data, the presence of
redshifted IPC absorption in \pgamma\ (9/38 CTTS) and \helium\
(18/38 CTTS) profiles implies that each of these lines probes
magnetospheric funnel flows. However, some unexpected
profile characteristics pose new challenges to our
understanding of the magnetospheric accretion process.

A hydrogen line such as \pgamma\ is expected to be formed
primarily in the accretion flow, and taken in aggregate, our
observations appear to support this expectation. The profiles
are single peaked with a range of line widths (\vbhi\ from -250
to -26 \kms), a tendency toward blueward centroids
(\vc\ from -150 to 29 \kms) and an IPC frequency of 24\%. However, the
extremes in linewidths and centroid velocities and their {\it
progression} with veiling are not predicted by existing models.

\begin{figure}
\epsscale{0.7}
\plotone{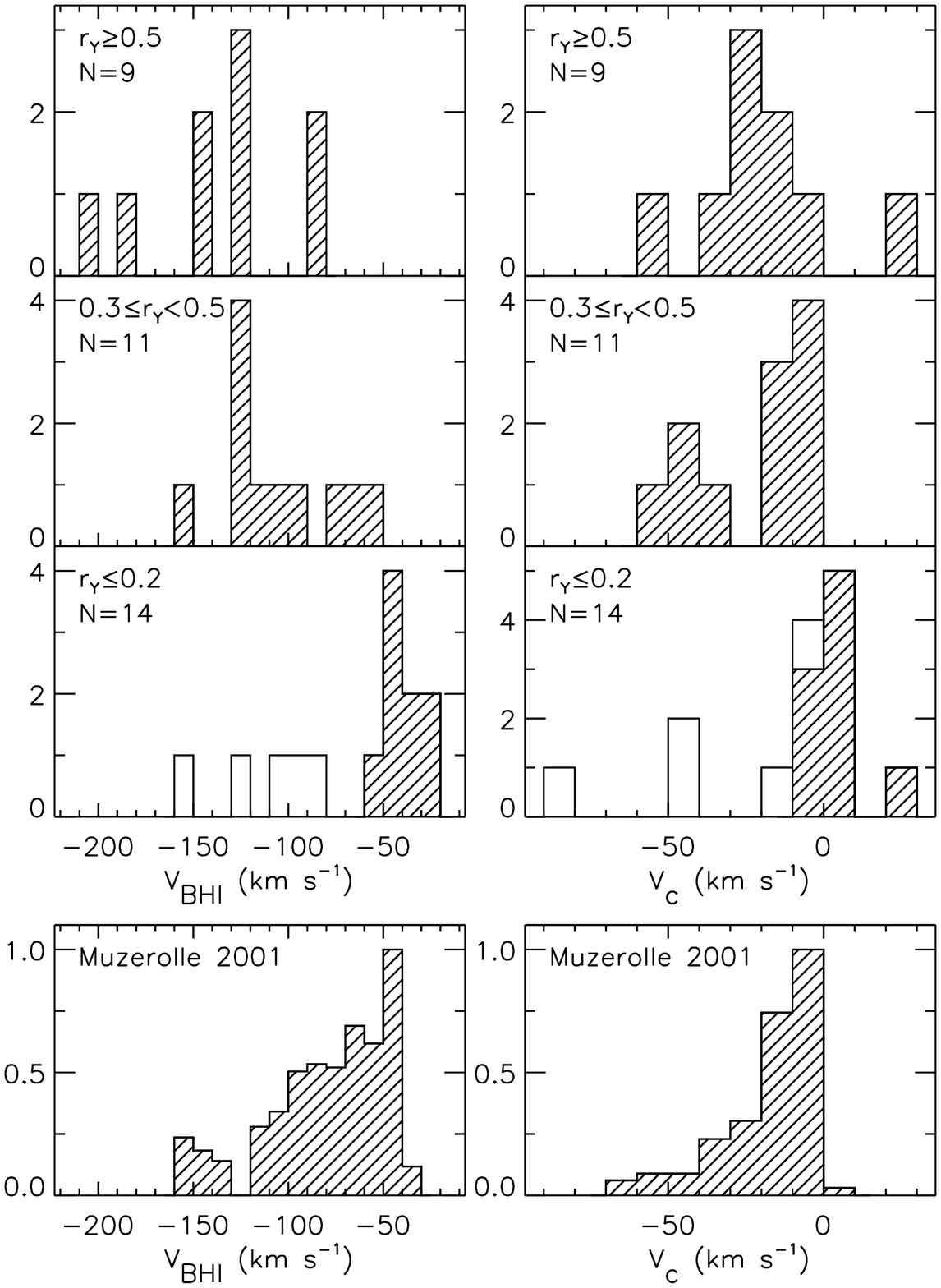}
\figcaption{ Comparision of histograms of kinematic parameters
\vbhi\ (left column) and \vc\ (right column) for observed and
theoretical Paschen line profiles. Observed values from \pgamma\
are separately shown in the first 3 rows for high, middle and
low 1\micron\ veiling. Theoretical values from online model
P$\beta$ profiles \citep{mch01} are in the bottom row.
The distribution for the low veiling group excludes 4 stars
with \pgamma\ profiles that bear no resemblance to the models
(BM And, SU Aur, CI Tau, UX Tau), making comparisons meaningless. We have
separately hatched the remaining 5 stars in the low veiling group with broad
\pgamma\ profiles to emphasize the trend toward decreasing line width and
centroid velocity with veiling.
The model histogram is normalized to unity
and weighted for a random distribution of viewing angles.
\label{f.pbeta}}
\end{figure}

To quantify the difference between observed and predicted profiles
we compare distributions of their kinematic parameters \vbhi\
and \vc\ in Figure~\ref{f.pbeta}. The kinematic parameters for
the data are separately plotted for \pgamma\ profiles in the
high, middle and low veiling groups described in Section 3.
Corresponding distributions for the theoretical profiles from
magnetospheric accretion models \citep{mch01} are shown in the bottom panel.
The theoretical parameters are measured from the online database
of P$\beta$ profiles predicted to arise under a variety of mass
accretion rates and disk truncation radii.\footnote{Available at
\url{http://cfa-www.harvard.edu/cfa/youngstars/models/magnetospheric\_models.html}.}
We selected only those profiles considered by Muzerolle to be
appropriate combinations of gas temperature and accretion rate,
which are $T=$~6000 or 7000~K at $\dot{M}=10^{-6}$~\msunyr,
$T=$~7000 or 8000~K at $\dot{M}=10^{-7}$~\msunyr, $T=$~8000 or
10,000~K at $\dot{M}=10^{-8}$~\msunyr, or $T=$~10,000 or
12,000~K at $\dot{M}=10^{-9}$~\msunyr. Following these
constraints, we extracted 128 of the available 272 stark
broadened model P$\beta$ profiles, covering disk truncation
radii from $\sim$2 to 6 R$_*$ seen from 4 viewing angles and
evaluated their \vbhi\ and \vc. To facilitate comparison between
the theoretical and observed distributions, we rescaled
 the \vbhi\ and \vc\ distributions measured from
the models to correspond to the appropriate weighting for a
random selection of orientation angles.

Although the model and observed profiles span similar ranges of
\vbhi\ and \vc\ for many of the stars, some observed profiles
show kinematic properties outside the range of the models. At
high veiling 2 of the 9 stars have \vbhi\ $\sim$ 50 \kms\ higher
than the models. More significantly, at low veiling many
stars have \vbhi\ $<$ 50 \kms\ and \vc\ $<$ 10 \kms, narrower
and more centered than predicted for magnetospheric accretion
flows. Moreover, as discussed in Section 3, these narrow and
centered \pgamma\ kinematic properties (3rd panel of
Figure~\ref{f.pgcat}) are actually from the narrow component of
profiles with a two-component (narrow and broad)
structure.

The \pgamma\ narrow component is reminiscent of a similar feature
in many metallic emission lines in CTTS, usually attributed to
formation in a stationary accretion shock at the footpoint of
the funnel flow \citep{bat96,joh97,ab00,naj00}. For example, the
FWHM of the narrow component at \heopt\ from the study of BEK,
which overlaps significantly with the stars studied here, is
$47\pm7$~\kms, in comparison to that for the average \pgamma\
value of $\sim$52 \kms. In CTTS metallic lines the proportional
contribution of narrow and broad component emission is sensitive
to the accretion rate, where the narrow component from the
accretion shock dominates over the broad
component predominantly in stars with low accretion rates.

\begin{figure}
\epsscale{1.0}
\plotone{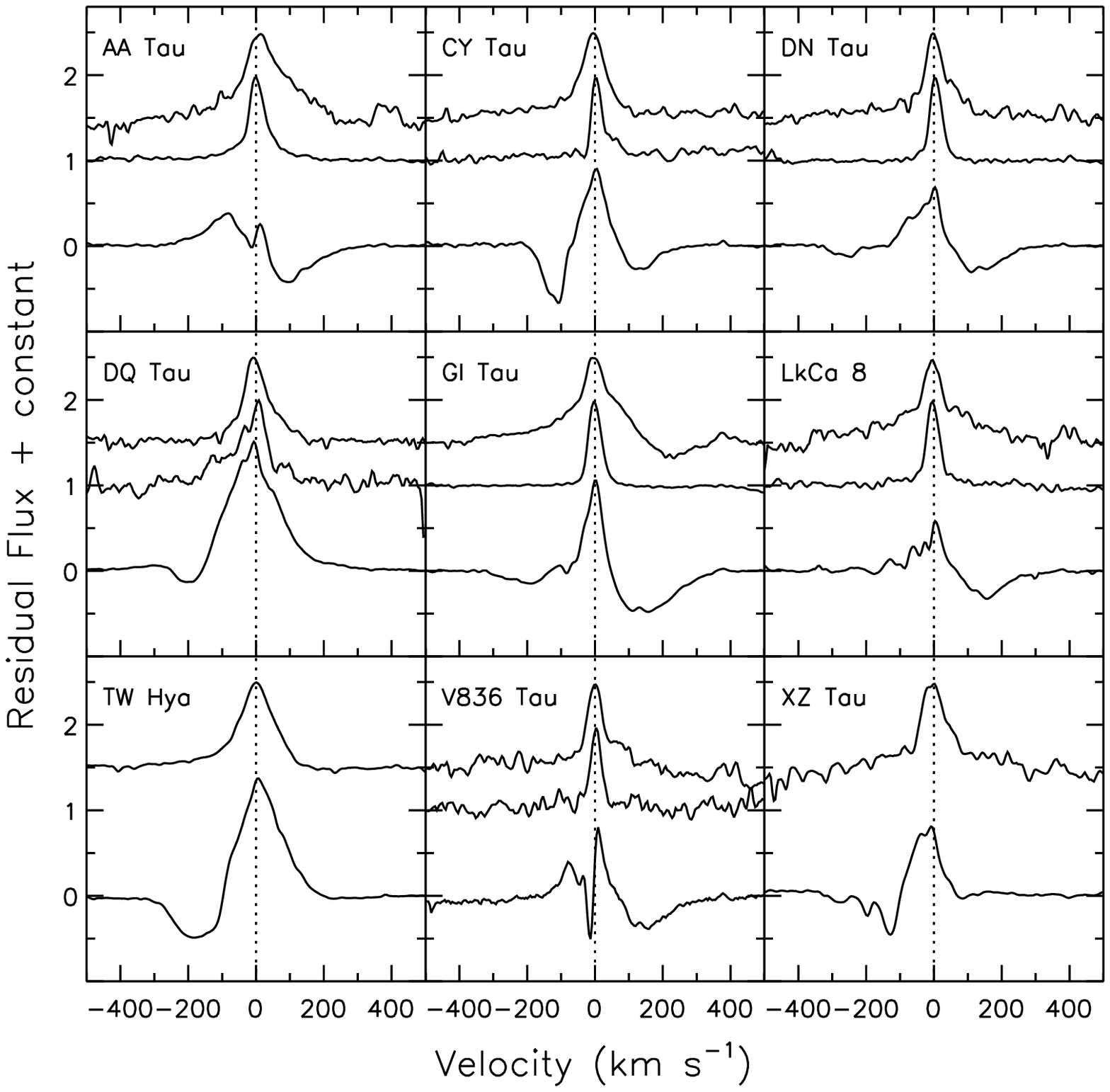}
\figcaption{ Comparison of \pgamma\ (top), \heopt\ (middle, from BEK),
and \helium\ (lower) profiles for  9 CTTS
with low veiling and narrow \pgamma\ lines.  \pgamma\ and \heopt\ are normalized
to their peak intensities to facilitate comparison of their kinematic
structure. The \helium\ profiles are in residual intensity units, except
for TW Hya which has been rescaled to half its actual peak intensity. No \heopt\
spectra are available for TW Hya or XZ Tau.
\label{f.nc_compare}}
\end{figure}

A comparison of \pgamma, \heopt\ (from BEK), and \helium\
profiles for the 9 stars with low 1~\micron\ veiling and
narrow \pgamma\ lines is shown in Figure~\ref{f.nc_compare}.
(Two were not in the BEK sample, but we include them
in the figure for completeness.)
Although the \heopt\ lines predate the 1~\micron\ lines by
almost a decade, it is apparent that all 3 lines in these stars
have relatively narrow emission, and for 7 stars for which
we have both \pgamma\ and \heopt\ the presence of a narrow
core near line center is apparent in both lines. In contrast \helium\
lines are considerably more complex due to the sensitivity of
this transition to absorption by infalling and outflowing gas.
However, in most cases the velocity extent of the blue
or red absorption at \helium\ appears to be matched by the
broad component emission at \pgamma\ or \heopt.

If \pgamma\ narrow component emission, which is only seen in low
veiling stars, arises in an accretion shock then the assumption
that hydrogen lines in low accretion rate stars primarily trace
magnetospheric columns is suspect. A related question is where
the accompanying \pgamma\ broad component is formed. In 3 of the
low veiling stars in Figure~\ref{f.nc_compare} there is an
asymmetric blue wing at \pgamma\ (AA Tau, GI Tau, TW Hya) with a
similar velocity extent to the blue absorption at \helium,
suggesting it may arise in the wind. A similar inference was
reached by Whelan et al. (2004) who found blue wings of hydrogen
lines in CTTS to be spatially extended, requiring formation in a
wind. The fact that the majority of the CTTS \pgamma\
profiles show a progression from broad blueshifted emission at
high veiling to narrow centered emission cores with broad bases
at low veiling could be explained if there is an increasing
contribution from wind emission at \pgamma\ as the veiling
increases. Or, if the lines are primarily formed in
magnetospheric accretion columns, this behavior needs to be
accounted for. In the models it is inclination that is the major
influence on line width and asymmetry, while the dominant effect
of accretion rate is on line luminosity, although when combined
with high temperatures line widths will also be enhanced \citep{mch01}. The
observed progression with veiling suggests that the paradigm of
attributing hydrogen line emission to magnetospheric accretion
might need to be readdressed, despite the considerable success
of previous comparisons of models and observations.

There is however little doubt that magnetospheric accretion is
present in CTTS, attested to by the presence of redshifted IPC
absorption in both hydrogen and metallic lines. Yet our
data poses challenges for IPC features as well, particularly
at \helium\ where the red absorption is often quite broad
(several hundred \kms) and deep (penetrating up to 50\% of the
continuum; see section 4 and Figures 4, 6, 10, 13).  This
kinematic structure suggests that magnetospheric infall
does not occur in a simple dipole geometry since the funnel flow
has a cross-section perpendicular to the line of sight that
narrows toward the star and becomes much smaller than the
projected stellar surface. In the beginning of the funnel flow
the speed of the gas also increases along a curved path that is
more perpendicular to the line of sight than along it, so it is
difficult to have the loci of points having the same
line-of-sight velocity component project a large cross-section
for a broad velocity range, as required for the observed
absorptions. Similar conclusions have also been reached by other
authors investigating the geometry of T Tauri magnetospheres via a
variety of techniques \citep{bouv, daou, jardine}.

We infer that although magnetospheric accretion along field lines
connecting the disk and star is common in CTTS, current models
of line formation in these flows, while contributing greatly to
our understanding of the star/disk interaction zone,
oversimplify the actual situation. Our data suggests that the
accretion shock can be a significant contributor to Paschen
emission in low accretion rate objects and that a wind may contribute
to the blue wings. Whether this calls into
question a magnetospheric accretion origin for \pgamma\ emission
in the majority of CTTS is unclear. Although
many of the observed \pgamma\ profiles have kinematic parameters
in line with those predicted, the models do not
account for the observed progression of \pgamma\ line widths and
centroid velocities with decreasing veiling shown in Figure~\ref{f.pgcat}.
Thus caution is in order when deriving
physical parameters of accretion flows on the sole basis of
hydrogen line profiles, since it is clear that even lines of
modest optical depth can have multiple sources of origin. In any
case, even with multiple emission components contributing to
\pgamma\ emission, each must increase in proportion to the disk
accretion rate in order to reproduce the correspondence between
line strength and $r_Y$ seen in Figure~\ref{f.eqwact}.

\subsection{Wind Contributions to \helium}

The \helium\ line appears to be a promising means
of studying the inner wind, given the high frequency (71\%) of CTTS with
subcontinuum blueshifted absorption. In comparison, the detection
frequency of subcontinuum blueshifted absorption at \halpha, a
traditional indicator of inner winds, is only 10\% (3 of 31) among CTTS
in a sample that closely overlaps the present one (see profiles
in \citetalias{BEK}). The most striking aspect of blueshifted
absorption among CTTS \helium\ profiles is their considerable
diversity -- in width, velocity, and penetration depth into the
continuum. Blue absorption widths have breadths as wide as 470
\kms\ to as narrow as 30 \kms. The narrow absorption features
can have blueshifts as high as 300 \kms\ to as low as tens of
\kms. Penetration depths
can be as low as 10\%, but exceed 50\% in 9 stars, and exceed 90\% in
two stars (DR Tau and TW Hya). This diversity indicates that the
innermost winds in accreting stars do not all have the same
characteristics, and that a range of wind conditions exist in
these stars. However, in all cases the winds must be accretion
powered, since there is no evidence for them in non-accreting
WTTS and the \helium\ line strength, in combined emission plus
absorption, is sensitive to the 1~\micron\ veiling. In this
section we highlight properties of the \helium\ blueshifted
absorption that speak directly to the wind launch region. We are
also undertaking a theoretical examination of profiles formed in
winds arising both from the star and the inner disk, using Monte
Carlo simulations to explore profiles formed via resonance
scattering and in-situ emission (Kwan et al., in preparation). A
comparison between observed and theoretical profiles will
elaborate on the points made here and demonstrate them more
rigorously.

Among the stars with ``P Cygni-like'' profiles (left column in
Figure~\ref{f.hecat} plus SVS 13 in Figure~\ref{f.class12}) five
have blue absorption which is {\it both} very broad and very
deep (SVS 13, DR Tau, AS 353A, HL Tau, and GW Ori). We argue
that these absorption features {\it require} formation in a wind
emerging from the star rather than the disk (see also Edwards et
al. 2003). In these cases the combination of the extreme line
breadth and depth reveals that all velocities in the wind, from
rest to terminal velocity, intercept the $1~\mu m$ continuum and
scatter a significant fraction of those photons. These geometric
constraints are satisfied when the source of the $1~\mu m$
continuum is the stellar surface and diverging wind streamlines
emerge radially away from the star, thereby absorbing continuum
radiation over the full acceleration region of the wind. In
contrast, for a disk wind geometry continuum photons from the
star will encounter a much narrower range of velocities as they
pass through the wind, which will be confined to nearly parallel
streamlines emerging at an angle to the disk surface. Even in an
x-wind \citep{shu94}, where wind streamlines open from a narrow
ring in the disk, the full range of velocities in an
accelerating wind would not occult the stellar surface. Central
to this interpretation is the assumption that the source of the
$1~\mu m$ continuum photons originates from the stellar surface.

This assumption for the location of the 1~\micron\ continuum
appears reasonable, as the continuum likely arises from the
photosphere itself or from accretion luminosity on the stellar
surface (see Section 7). Even for the largest continuum excesses
we observe, $r_Y = 2$, a third of the continuum is contributed
by the photosphere. The 5 stars with the most broad and deep
blue absorption of the 1~\micron\ continuum, and thus the
strongest evidence for winds moving radially away from the star,
have $r_Y$ values of 2 (DR Tau), 1.8 (AS 353A), 1 (HL Tau) and 0
.3 (SVS 13 and GW Ori), demonstrating that similar absorption
features are seen in stars with a range of continuum excesses
and photospheric contributions. We refer to this wind geometry
as a {\it stellar wind}, although these winds are clearly
deriving their energy from processes related to disk accretion.

The above argument for stellar winds is based on the structure of
the blue absorption feature in 5 stars. This interpretation is
also consistent with their full helium profiles, which resemble
line formation in a stellar wind under conditions of simple
resonance scattering. Such conditions would normally result in
classic P Cygni profiles with comparable emission and absorption
contributions, as seen for AS353 A, HL Tau, and GW Ori. However,
if the accretion disk extends close enough to the star to occult the
receding flow and suppress the red emission, then a stellar wind
can also account for the full profiles of DR Tau and SVS 13
where \helium\ is predominantly in absorption. This would result
if the stellar wind starts beyond the disk truncation radius, which
would be favored, for example, under conditions of high disk accretion
rate, resulting in smaller truncation radii \citep{k91}.

For the remaining CTTS interpretation of the \helium\ profiles is
less definitive, although stellar winds are still good candidates for the
profiles in the P Cygni group, such as DG Tau,
DF Tau, DO Tau, and TW Hya. In these stars the blue
absorption is narrow but very blueshifted and emission is the
dominant contributor to the profile. Such profiles could be
explained if resonance scattering plus an additional source of
in-situ emission is present in a stellar wind, where prominent
emission would also partially fill in the absorption feature, as
observed.

\begin{figure}
\plotone{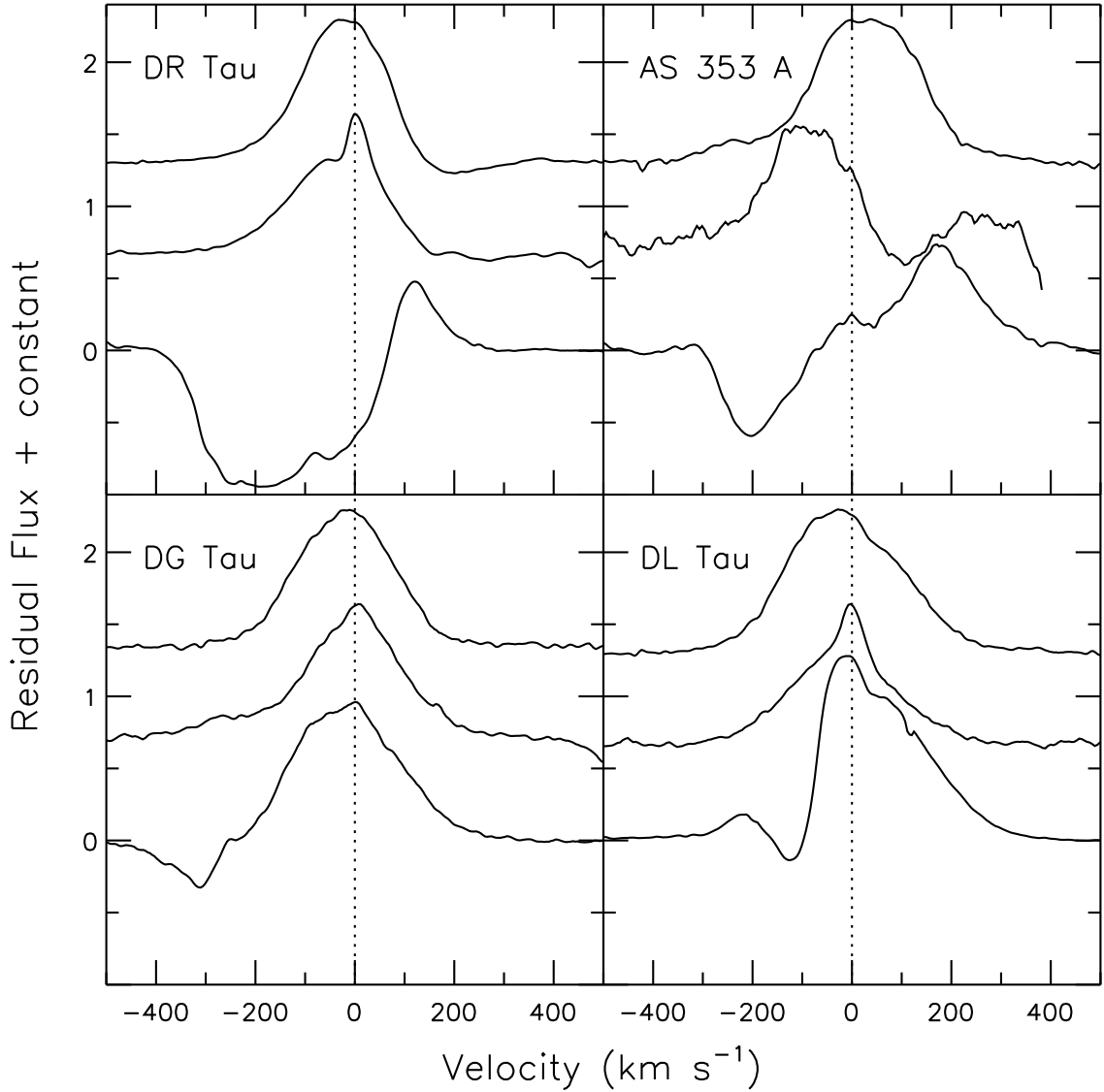}
\figcaption{ Comparison of \pgamma\ (top), \heopt\ (middle, from BEK),
and \helium\ (lower) profiles for 4 CTTS
with high veiling and broad \pgamma\ lines.  \pgamma\ and \heopt\ are normalized
to their peak intensities to facilitate comparison of their kinematic
structure. The \helium\ profiles are in residual intensity units.
\label{f.bc_compare}}
\end{figure}

The possibility raised in the previous section, that the wind may
contribute to broad component \pgamma\ emission, gains
additional support when \pgamma\ profiles from stars with high
veiling are compared with helium lines. In
Figure~\ref{f.bc_compare} we present a figure analogous to that
of Figure~\ref{f.nc_compare} showing \pgamma, \heopt, and
\helium\ profiles, but in this case for 4 stars with large
1~\micron\ veiling and P Cygni-like \helium\ profiles. In contrast
to the narrow emission seen at all 3 lines in the low veiling stars,
here most of the lines show broad emission (or in some cases,
absorption). Similarly broad, blueshifted emission is seen for
both \pgamma\ and the non-simultaneous \heopt\ lines for 3 of these stars
(DR Tau, DG Tau and DL Tau). In BEK we argued that the broad
blueshifted component of  \heopt\ arose primarily in a wind, which
Figure~\ref{f.bc_compare} suggests may also be contributing to
\pgamma\ emission. If so, this effect may be an important
contributor to the progression of \pgamma\ profile morphology
described in Sections 3 and 6.1. The 4th star, AS 353A, shows a
more confusing relation between \pgamma\ and the non-simultaneous
 \heopt\ line, where it
appears as if the blue side of \pgamma\ and the red side of
\heopt\ may be influenced by absorption.

In addition to the P Cygni profiles with blueshifted absorption
likely arising from accretion powered
stellar winds, other CTTS \helium\ profiles have blue absorption that is both
narrow and less blueshifted, e.g. DS Tau, UY Aur, V836 Tau. Here
disk winds provide a natural explanation for the blue
absorption, where stellar photons are intercepted over a
narrower range of velocities as they cross parallel wind
streamlines emerging from the disk. The magnitude of the
blueshift will depend on the relative orientations of the disk
and the line of sight, where blueshifts would be smallest for
more edge-on systems. There are also a number of stars where the
interpretation of the blueshifted absorption is ambiguous, and on
the basis of the \helium\ profile alone, either a stellar or
disk wind might be invoked. There are also 4 CTTS with no
subcontinuum blue absorption and 2 with deep, centered
absorptions, reminding us that the star-disk interface region is
complex and not uniform among all accreting stars.

We consider it likely that both disk and stellar winds are
present in most CTTS, but that a combination of factors
including inclination, disk geometry, accretion/outflow rates,
magnetospheres, and accretion shock properties will all
contribute to the formation of an observed profile. If both
types of winds are present, then the stellar winds probably
emerge predominantly from high latitudes. If so, the signature
of stellar winds would be favored in stars seen more pole-on,
while disk wind signatures would be favored in stars seen more
edge-on. Unfortunately, inclination estimates for most CTTS are
not well constrained, and accretion rates are variable, so we
have not tried to sort out these effects for our profiles at the
present time. Although \helium\ profile morphologies do not have
a one-to-one correspondence with the 1~\micron\ veiling, the
fact that good examples of stellar wind signatures are found for
stars thought to have high accretion rates and well formed jets
(e.g. SVS 13, DR Tau, AS 353A, HL Tau, DG Tau) clarifies that
the stellar wind component can be important in stars with
significant accretion activity.

\section{ORIGIN OF 1~\micron\ VEILING}

Continuum emission excesses that veil photospheric features in
CTTS spectra are known to come from at least 2 different
sources. One is an optical/ultraviolet excess shortward of
0.5~\micron\ that is used to determine disk accretion rates. The
spectral energy distribution of the excess emission from 0.32 to
0.52~\micron\ (Valenti, Basri, \& Johns 1993; Gullbring et al.\
1998) is very well described by theoretical models from
accretion shocks formed as magnetically channeled material falls
from the disk truncation radius to the stellar surface along
magnetospheric field lines (Calvet \& Gullbring 1998). In the
models the observed continuum emission is attributed to
optically thick Paschen continua arising in the heated
photosphere (6000-8000 K) below the shock plus optically thin
emission shortward of the Balmer limit from preshock and
postshock gas. There are indications that the emission excess
longward of $0.5 \mu m$ does not steadily decline with
wavelength as expected from the accretion shock models, but
this has not been systematically studied since it would require
simultaneous observational coverage over a wide wavelength region.
However, distributions
of veiling among similar ensembles of CTTS indicate that it
appears to be relatively constant from $0.5~\mu m$ to $0.8 ~\mu m$ (Basri
\& Batalha 1990; White and Hillenbrand 2004), suggesting that
there is an additional contribution to emission excess. This leads to
large uncertainties in bolometric corrections and in corresponding
accretion luminosities and mass accretion rates from veilings
measured in the red.

A second source of excess continuum emission
recognized in accretion disk systems is found at near infrared
wavelengths from 2 to 5~\micron\ \citep{folha99,jkv01}. This
emission excess is well characterized by a single temperature
black body with $T\sim1400$~K and has been successfully modelled
as arising in a raised dusty rim at the dust sublimation radius
in the disk \citep{muz03}. The amount of the excess is proportional
to the accretion luminosity, as the dust is heated by radiation from
both the photosphere and the accretion shock. The location of the
dust sublimation radius is about 0.07 AU for a K7/MO star but can be
expanded to 0.15 AU when accretion luminosities are high
( Eisner et al. 2005).

\begin{figure}
\plotone{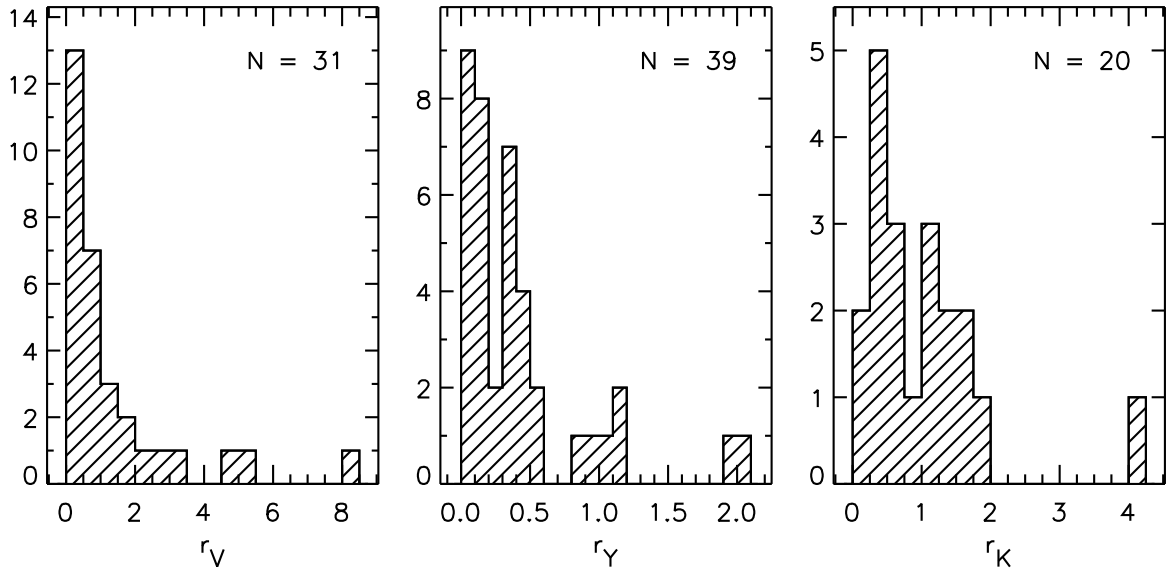}
\figcaption{ Comparison of veiling distributions at V, Y and K. Only stars from the current
survey are included, but the non-simultaneous data for V and K from the literature
does not include all the stars in our survey.
\label{f.veil}}
\end{figure}

Our NIRSPEC data provides the first reconnaissance of veiling at
$1~\mu m$. This wavelength regime is ideally situated to provide
additional insight into the behavior of continuum excess in CTTS
since it lies between the regions where 6000-8000 K shocked gas and
1400 K warm dust are dominant contributors. However, this also
corresponds to the region where the photospheric emission peaks,
which will result in smaller veiling, defined as an excess
relative to photospheric flux, for comparable emission fluxes.
This is in fact seen, as shown in Figure~\ref{f.veil}, which
compares distributions of $r_V$, $r_Y$, and $r_K$ for CTTS,
using veilings from the literature for the stars in our NIRSPEC
survey. It is seen that maximum values for $r_Y$ are only
$\sim2$, compared to maxima of 8 for $r_V$
and 4 for $r_K$. Similarly 20\% of stars with
optical excesses do not have measurable
veilings at 1~\micron. (Veiling values and their references are
compiled in Table~\ref{t.veiling}). However, as we show below,
the 1~\micron\ veiling on average exceeds that expected from
either a 6000-8000 K accretion shock or 1400 K dust, and leads us to
confirm that an additional source of excess emission is
present in CTTS systems.

\begin{figure}
\plotone{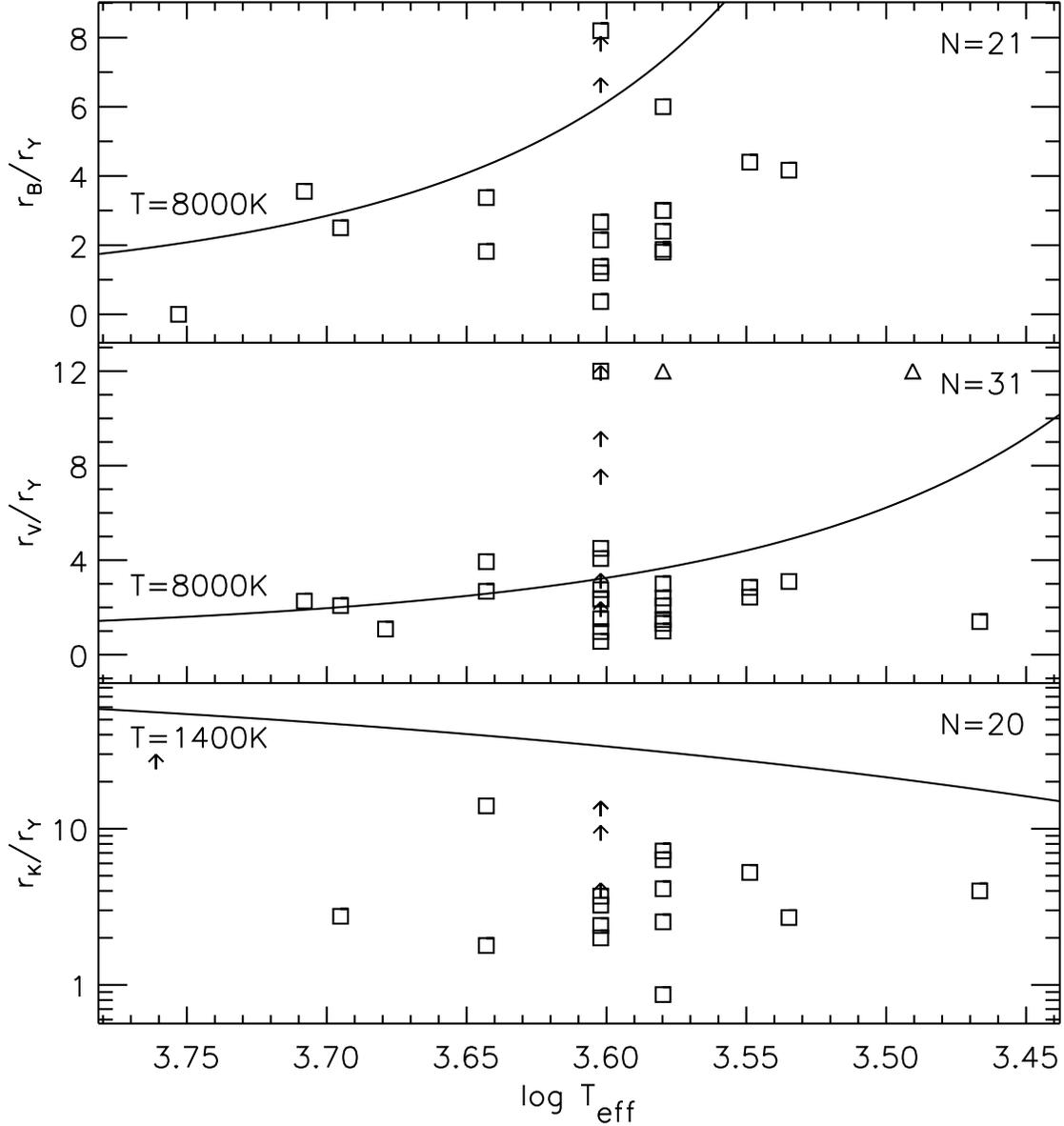}
\figcaption{Ratios of $B$, $V$, and $K$ band veiling to $Y$ band
veiling plotted as a function of effective temperature. Stars
with measured veilings at each pair of wavelengths are plotted
as open squares using mean values given in
Table~\ref{t.veiling}. Two stars in the $r_V/r_Y$ plot have
ratios too high to fit in the figure ( 15.7 for DO Tau and 29.0
for DD Tau) and are shown with triangles. To include objects
with non-detections at $r_Y$, we set $r_Y$ to an upper limit of
$r_Y=0.025$. These are shown as open arrows, representing lower
limits on the ratio. The solid line shows the predicted ratio
for a photospheric $T_{\rm{eff}}$ with an excess blackbody
source of 8000~K (top two panels) or 1400~K (bottom panel). Note
that the ratios are scaled linearly in the first two panels and
logarithmically in the lower one.
\label{f.rtemp}}
\end{figure}

The magnitude of the discrepancy is shown in
Figure~\ref{f.rtemp}, which uses veiling ratios,
$r_\lambda/r_Y$, to illustrate how, on average, veilings at 1~\micron\
exceed those predicted from either of the two suspected sources of
continuum excess emission. The veiling ratios are
plotted as a function of $T_{\rm{eff}}$ in order to account for
the differing proportions of photospheric contributions to the
measured veiling. We also include predicted
veiling ratios that would result either from 8000K accretion
shocks ($r_B/r_Y$ and $r_V/r_Y$) or 1400 K warm dust ($r_K/r_Y$).
Although the CTTS ratios are compiled from non-simultaneous
veiling measurements, the ensemble of veiling ratios from many
stars consistently shows $r_Y$ to be larger than expected from
either accretion shocks or warm dust.

For example, the expected veiling ratio $r_K$/$r_Y$ is about 20 for warm 1400
K dust around a K7/M0 star. Thus the observed maximum $r_K$
of $4$, would only generate a $1~\mu m$ veiling of $r_Y$ =0.2,
considerably smaller than the observed maximum of 2. Similarly, the
observed distribution of $r_K$/$r_Y$ ratios indicates that on average $r_Y$ is
an order of magnitude larger than would arise from the dust.
 
The corresponding situation for an 8000 K accretion shock
predicts a veiling ratio $r_B/r_Y$ of about 7 for a K7/M0 star.
Thus the maximum $r_B$ of 5 would correspond to an $r_Y$ of 0.7
in contrast to the observed maximum of 2. This is consistent
among the group of CTTS, where the observed ratios again suggest
$r_Y$ is, on average, several times larger than predicted. A
similar conclusion can be drawn from the ratio $r_V/r_Y$,
although the average discrepancy here is not as large. This is
because by 5700 \AA\ the excess emission is already higher than
expected relative to the B-band for an 8000 K accretion shock.

In summary, there is persuasive evidence suggesting that an additional
source of continuum emission excess other than 6000-8000 K accretion shocks
and 1400 K warm dust is present in accreting T Tauri stars. A simple
possibility is that accretion shocks are not uniform in temperature, but
include cooler components than modelled to date (White and Hillenbrand 2004).
Another is radiation emerging from the region where the rotational energy
of the inner disk is dissipated, which would generate lower temperatures
than the accretion shocks. Determining the
source of this emission and its importance in accreting systems
will require simultaneous measurements over a very broad range of
wavelengths.

\section{CONCLUSIONS}

{\it Our key empirical findings from NIRSPEC 1~\micron\ spectra are:}

Veiling at 1~\micron\ ranges from $r_Y$ = 0 to 2 for a sample of
39 CTTS whose veiling at 5700 \AA\ ($r_V$ = 0.1 - 8)
corresponds to two orders of magnitude in disk accretion rates.
Approximately 20\% of the CTTS (8/39) showed no detectable
continuum excess at 1~\micron\, ($r_Y < 0.05$).
For the 3 Class I sources we observed, 2 have 1 \micron\ veiling.
In the CTTS the 1~\micron\ veiling is on average larger than
expected from recognized sources of continuum
excess in accreting stars, either 6000-8000 K accretion shocks on the
stellar surface or 1400 K puffed rims at the dust sublimation
radius in the disk.

The \pgamma\ line is in emission in 38/39 CTTS with equivalent
widths ranging from 0.6 to 18 \AA and in all 3 Class I sources.
Profiles have central peaks,
none show blueshifted absorption, and 24\% of the CTTS show redshifted absorption
below the continuum. There is good correlation between \pgamma\
equivalent width and veiling and the majority of stars also show
a relation between \pgamma\ profile morphology and $r_Y$, with
broader more blueshifted lines at high veiling. Among the 18
stars with $r_Y \le 0.2$, half show \pgamma\ profiles with a two
component structure comprised of narrow centered cores ($FWHM
\sim 50$ \kms) and a broad base ($FWHM \sim 170$ \kms).

The \helium\ line, found in 38/39 CTTS and all 3 class I
sources, displays a more complex profile morphology. Its resonance
scattering properties make it extraordinarily sensitive to
outflowing and infalling gas in the
vicinity of star, with 89\% of the CTTS profiles showing
absorption below the continuum. Blueshifted subcontinuum
absorption from a wind is found in 71\%
of the CTTS profiles, redshifted subcontinuum absorption from
magnetospheric infall in 47\%, and centered
subcontinuum absorption in 5\%.
Blueshifted absorption is found at all veiling levels and
displays considerable diversity in its velocity structure, with
widths ranging from 30- 470 \kms and depths penetrating up to
95\% of the continuum.
Redshifted absorption is more prevalent among stars with low
veiling and when present can have widths of several hundred \kms
and penetrate up to 50\% of the continuum.

{\it Our key interpretive findings from NIRSPEC 1~\micron\ spectra are:}

We attribute the diversity of kinematic structure in \helium\
subcontinuum blueshifted absorption to the presence of two
genres of winds being launched from the star-disk interface
region. Some CTTS have profiles requiring a wind moving radially
away from the star while others are more readily explained as
arising in winds emanating from the inner disk. There are also
stars where it is not obvious which genre of wind is dominant.
Both winds must be accretion powered since neither is seen in
non-accreting WTTS and among the CTTS helium strength correlates
with veiling. Stellar wind signatures are found for a range of
disk accretion rates, including objects of high accretion rate
and well developed jets and HH outflows. We consider it likely
that both disk and stellar winds are present in all CTTS, but
that a combination of factors, including inclination to the line
of sight, disk geometry, accretion/outflow rates, magnetospheres,
and accretion shock properties determine which
genre of wind is the primary contributor to the \helium\ profile
in a particular star.

The high frequency of redshifted subcontinuum absorption in half
of the \helium\ profiles and a quarter of the \pgamma\ profiles
attests to the presence of magnetospheric accretion flows.
However, the breadth and depth of this feature at \helium\
suggests that complex field topologies rather than simple dipole fields
probably link accretion columns from the disk to the star, an
idea for which there is growing observational support \citep{bouv}.
Surprisingly, several aspects of \pgamma\ profile morphology
suggest this line may not be solely formed in magnetospheric
infall, but may also have important contributions from accretion shocks
and winds.

In sum, the combination of \helium\ and \pgamma\ lines in the 1~
\micron\ region offer rich new probes of the relation between
accretion and outflow in young stars. This paper presents the
first empirical overview of these profiles among CTTS and
demonstrates the unique power of the \helium\ line to probe the
wind launch region in accreting stars. A companion paper will
present theoretical profiles for lines formed via scattering and
in-situ emission in winds arising from stars and from disks. The
finding that at least some of the highest accretion rate CTTS
show signatures for accretion powered stellar winds
clarifies that stellar
winds must be an important component in accreting systems. This
allows for the possibility that magnetized stellar winds rather
than magnetospheric disk locking may be the means by which
accreting stars spin down to rotational velocities well below
break-up \citep{mat05,rb05,love05}.

Establishing mass outflow rates
from \helium\ will require good determination of physical
conditions in the helium wind, which in turn will require
simultaneous spectra of both \helium\ and \heopt\ to establish
the conditions for helium excitation and ionization. Until then
the magnitude of the contribution of winds traced by helium,
either stellar or inner disk winds, to extended mass outflows
and jets and to angular momentum loss from the star or
disk remains uncertain.

{\it Acknowledgements}:
NASA grant NAG5-12996 issued through the Office of
Space Science provides support for this project.
Thanks to Andrea Dupree for conversations and collaboration
in an earlier phase of this work, to Russel White and Catrina
Hamilton for IDL scripts and to
the queue observers for the Keck-Gemini NIRSPEC
program, Tom Geballe and Marianne Takamiya for acquiring the 2001
data, and to an anonymous referee for helpful comments.

\clearpage

\begin{deluxetable}{lcccccccc}
\tablecaption{T Tauri Sample \label{t.sample}}
\tabletypesize{\footnotesize}
\setlength{\tabcolsep}{0.04in}
\tablecolumns{9}
\tablewidth{0pt}

\tablehead{ \colhead{Object} & \colhead{Sp Type} & \colhead{HJD} &
\colhead{$r_{V}$} & \colhead{$\log \dot{M}_{\rm{acc}}$} &
\colhead{Ref} & \colhead{$\log \dot{M}_{\rm{wind}}$} & \colhead{Ref} &
\colhead{$\dot{M}_{\rm{wind}}/\dot{M}_{\rm{acc}}$} \\ \colhead{(1)} &
\colhead{(2)} & \colhead{(3)} & \colhead{(4)} & \colhead{(5)} &
\colhead{(6)} & \colhead{(7)} & \colhead{(8)} & \colhead{(9)}}

\startdata
\cutinhead{CTTS}
AA Tau\dotfill   & M0 & 605.0,606.9* & 0.15-0.63 & -8.5 & 1 & -9.1 & 2
& 0.25 \\
AS 353 A\dotfill & K5 & 605.7,606.7* & 5.10 & -5.4 & 2 & -7.6 & 2 &
6.3x10$^{-3}$ \\
BM And\dotfill   & K5 & 604.8 & \nod & $>$-9* & 3 & \nod & \nod & \nod \\
BP Tau\dotfill   & K7 & 605.9 & 0.41-0.63 & -7.5 & 1 & (-9.7) & 2 &
(6.3x10$^{-3}$) \\
CI Tau\dotfill   & K7 & 605.9 & 0.39-0.54 & -6.8 & 2 & -8.9 & 2 & 7.9x10$^{-3}$
\\
CW Tau\dotfill   & K3 & 604.8,607.0* & 1.70 & -6.0 & 2 & -7.1 & 2 &
7.9x10$^{-2}$ \\
CY Tau\dotfill   & K7 & 606.8 & 1.20 & -8.1 & 1 & -10.0 & 2 & 1.3x10$^{-2}$ \\
DD Tau\dotfill   & M4 & 604.9 & 2.90 & -6.3 & 2 & -8.5 & 2 & 6.3x10$^{-3}$ \\
DE Tau\dotfill   & M1.5 & 606.0 & 0.57 & -7.6 & 1 & -9.2 & 2 & 2.5x10$^{-2}$ \\
DF Tau\dotfill   & M2 & 605.0 & 0.52-1.60 & -6.8 & 1 & -8.3 & 2 & 3.2x10$^{-2}$
\\
DG Tau\dotfill   & K5 & 235.9*,605.8 & 2.00-3.60 & -5.7 & 2 & -6.5 & 2
& 0.16 \\
DK Tau\dotfill   & K7 & 604.9,606.9* & 0.49 & -7.4 & 1 & -8.5 & 2 &
7.9x10$^{-2}$ \\
DL Tau\dotfill   & M0 & 605.0 & 1.10-2.40 & -6.7 & 2 & -8.9 & 2 & 6.3x10$^{-3}$
\\
DN Tau\dotfill   & K7 & 606.0 & 0.08 & -8.5 & 1 & -9.4 & 2 & 0.13 \\
DO Tau\dotfill   & M0 & 604.8 & 4.70 & -6.8 & 1 & -7.5 & 2 & 0.20 \\
DQ Tau\dotfill   & K7 & 606.0 & 0.17-0.18 & -9.2 & 1 & -8.7 & 2 & 3.2 \\
DR Tau\dotfill   & K7 & 605.0*,606.0,606.9 & 6.40-10.0 & -5.1 & 2 &
-8.6 & 2 & 3.1x10$^{-4}$ \\
DS Tau\dotfill   & K2 & 605.9 & 0.96 & -7.9 & 1 & (-9.3) & 2 & (4.0x10$^{-2}$) \\
FP Tau\dotfill   & M5 & 605.0 & 0.08-0.18 & -7.7 & 2 & (-10.7) & 2 &
(1.0x10$^{-3}$) \\
GG Tau\dotfill   & M0 & 605.0 & 0.14-0.50 & -7.8 & 1 & -9.1 & 2 & 5.0x10$^{-2}$
\\
GI Tau\dotfill   & M0 & 606.0 & 0.24 & -8.0 & 1 & -9.4 & 2 & 4.0x10$^{-2}$ \\
GK Tau\dotfill   & K7 & 606.0 & 0.23 & -8.2 & 1 & -9.2 & 2 & 0.10 \\
GM Aur\dotfill   & K7 & 605.1 & 0.21-0.22 & -8.0 & 1 & (-10.0) & 2 &
(1.0x10$^{-2}$) \\
GW Ori\dotfill   & G5 & 606.1 & \nod & -6.4 & 4 & \nod & \nod & \nod \\
HK Tau\dotfill   & K7 & 606.1 & 1.10 & -6.5 & 2 & -8.8 & 2 & 5.0x10$^{-3}$ \\
HL Tau\dotfill   & K7-M2 & 235.9,605.8* & \nod & -5.4 & 5 & -6.8 & 6 &
4.0x10$^{-2}$ \\
HN Tau\dotfill   & K7 & 604.8 & 0.76 & -8.9 & 1 & -8.1 & 2 & 6.3 \\
LkCa 8\dotfill   & M0 & 604.9 & 0.05-0.24 & -9.1 & 1 & -9.6 & 2 & 0.32
\\
RW Aur A\dotfill & K1 & 605.1 & 1.70-2.00 & -5.8 & 2 & -7.6 & 2 & 1.6x10$^{-2}$
\\
RW Aur B\dotfill & K3 & 605.1 & \nod & -7.3 & 6 & -8.0 & 6 & 0.20 \\
SU Aur\dotfill   & G2 & 607.0 & \nod & -8.0 & 4 & \nod & \nod & \nod \\
TW Hya\dotfill   & K7 & 605.2*,606.1 & \nod & -9.3 & 7 & \nod & \nod &
\nod \\
UX Tau\dotfill   & K2 & 607.0 & \nod & \nod & \nod & \nod & \nod &
\nod \\
UY Aur\dotfill   & M0 & 605.0,607.0* & 0.20-1.30 & -7.2 & 1 & -8.2 & 2
& 0.10 \\
UZ Tau E\dotfill & M1.5 & 605.9 & 0.73 & -5.7 & 2 & -7.6 & 2 & 1.3x10$^{-2}$ \\
UZ Tau W\dotfill & M3 & 605.9 & \nod & -8.0 & 8 & \nod & \nod & \nod \\
V836 Tau\dotfill & K7 & 606.0 & 0.00-0.06 & -8.2 & 2 & (-10.1) & 2 &
(1.3x10$^{-2}$) \\
XZ Tau\dotfill   & M3 & 605.8 & \nod & -8.1 & 9 & \nod & \nod & \nod \\
YY Ori\dotfill   & K7 & 607.1 & 1.80 & -5.5 & 2 & (-9.6) & 2 & (7.9x10$^{-5}$) \\
\cutinhead{WTTS}
LkCa 4\dotfill   & K7 & 607.0 & 0.0 & \nod & 2 & \nod & \nod & \nod \\
TWA 3\dotfill    & M3 & 606.1 & \nod & -10.3 & 7 & \nod & \nod & \nod \\
TWA 14\dotfill   & M0 & 606.2 & \nod & -10.3 & 10 & \nod & \nod &
\nod \\
V819 Tau\dotfill & K7 & 235.9 & 0.0 & \nod & 2 & \nod & \nod & \nod \\
V826 Tau\dotfill & K7 & 604.9 & 0.0 & \nod & 2 & \nod & \nod & \nod \\
V827 Tau\dotfill & K7 & 606.0 & 0.0 & \nod & 2 & \nod & \nod & \nod \\
\enddata

\tablecomments{Col 2: Spectral types from \citetalias{HEG}, \citet{HBC},
\citet{reid03}; Col 3: Heliocentric Julian Date (2,452,000 +);
for multiple observations an asterisk indicates membership in the
reference sample; Col 4: Veiling at 5700~\AA\ from \citetalias{HEG}; Col 5:
mass accretion rate in \msunyr; Col 7:
mass loss rate in \msunyr; parentheses indicate an
upper limit; Col 9: Ratio of the mass loss rate to the mass accretion
rate}

\tablerefs{(1) \citealt{GHBC}; (2) \citetalias{HEG}; (3)
\citealt{gue93}; (4) \citealt{gul00}; (5) \citealt{cal94}; (6)
\citealt{wh04}; (7) \citealt{mcbh}; (8) \citealt{hk03}; (9)
\citealt{vbj}; (10) \citealt{muz01}.}
\end{deluxetable}

\clearpage

\begin{deluxetable}{lccccccccccccc}
\tablecaption{CTTS Veilings and Profiles \label{t.kin}}
\rotate
\tabletypesize{\scriptsize}
\setlength{\tabcolsep}{0.04in}
\tablecolumns{14}
\tablewidth{0pt}
\tablehead{\multicolumn{2}{c}{} & \multicolumn{7}{c}{\pgamma} &
\multicolumn{5}{c}{He I $\lambda$10830} \\ \multicolumn{2}{c}{} &
\multicolumn{7}{c}{------------------------------------------------------------------------------------------}
&
\multicolumn{5}{c}{------------------------------------------------------------------------------}
\\ & & \colhead{$V_{\rm{bwing}}$} & \colhead{\vbhi} & \colhead{\vc} &
\colhead{Peak} & \colhead{Em $W_\lambda$} & \colhead{Ab $W_\lambda$} &
\colhead{Ab} & \colhead{$V_{\rm{bwing}}$} & \colhead{$V_{\rm{rwing}}$}
& \colhead{Em $W_\lambda$} & \colhead{Ab $W_\lambda$} & \colhead{Ab} \\
\colhead{Object} & \colhead{$r_Y$} & \colhead{(\kms)} & \colhead{(\kms)} & \colhead{(\kms)} &
\colhead{Intensity} & \colhead{(\AA)} &  \colhead{(\AA)} & \colhead{Type} & \colhead{(\kms)} &
\colhead{(\kms)} & \colhead{(\AA)} & \colhead{(\AA)} &  \colhead{Type} \\
\colhead{(1)} & \colhead{(2)} & \colhead{(3)} &\colhead{(4)} & \colhead{(5)} &
\colhead{(6)} & \colhead{(7)} & \colhead{(8)} & \colhead{(9)} & \colhead{(10)} &
\colhead{(11)} & \colhead{(12)} & \colhead{(13)} & \colhead{(14)} }
\startdata
\cutinhead{CTTS: High 1~\micron\ Veiling}
DR Tau\dotfill   & 2.0* & -400 & -122 & -25 & 1.64 & 12.6 & 0.4 & r & -400 & 400 & 1.6 & 10.7 & b  \\
                 & 2.0\phm{*} & -400 & -111 & -21 & 1.80 & 12.3 & 0.4 & r & -450 & 400 & 1.2 & 11.4 & b  \\
                 & 2.0\phm{*} & -350 & -108 & -17 & 1.93 & 13.7 & 0.3 & r & -450 & 400 & 0.1 & 12.5 & b,r  \\
AS 353 A\dotfill & 2.0\phm{*} & -400 & -108 & 18 & 1.54 & 17.7 & \nod & \nod & -300 & 500 & 4.8 & 2.2 & b  \\
                 & 1.8* & -350 & -90 & 29 & 1.40 & 14.6 & \nod & \nod & -400 & 500 & 5.4 & 2.8 & b  \\
CW Tau\dotfill   & 1.0\phm{*} & -400 & -167 & -46 & 0.96 & 10.2 & \nod & \nod & -450 & 450 & 14.0 & \nod & \nod  \\
                 & 1.3* & -350 & -82 & -23 & 1.71 & 11.8 & \nod & \nod & -400 & 400 & 16.4 & \nod & \nod  \\
DL Tau\dotfill   & 1.1\phm{*} & -350 & -149 & -18 & 1.41 & 14.5 & \nod & \nod & -500 & 450 & 21.4 & 0.4 & b  \\
HL Tau\dotfill   & 1.0\phm{*} & -250 & -144 & -15 & 0.89 & 9.5 & \nod & \nod & -200 & 350 & 2.5 & 4.0 & b  \\
                 & 1.0* & -300 & -122 & -2 & 1.19 & 10.4 & \nod & \nod & -400 & 400 & 4.5 & 3.8 & b  \\
DG Tau\dotfill   & 0.9* & -350 & -141 & -20 & 0.93 & 8.7 & \nod & \nod & -500 & 350 & 8.7 & 1.2 & b  \\
                 & 0.7\phm{*} & -300 & -138 & -25 & 0.77 & 6.6 & \nod & \nod & -350 & 350 & 6.8 & 0.1 & b  \\
RW Aur A\dotfill & 0.9\phm{*} & -400 & -186 & -52 & 1.20 & 11.9 & 0.2 & r & -350 & 300 & 15.0 & \nod & \nod  \\
DK Tau\dotfill   & 0.5* & -350 & -124 & -35 & 0.49 & 3.4 & \nod & \nod & -500 & 400 & 0.2 & 5.8 & b,r  \\
                 & 0.5\phm{*} & -350 & -83 & 4 & 0.35 & 2.7 & \nod & \nod & -500 & 400 & \nod & 6.7 & b,r  \\
HN Tau\dotfill   & 0.5\phm{*} & -400 & -201 & -23 & 0.69 & 9.6 & \nod& \nod & -500 & 500 & 30.1 & \nod & \nod  \\
\tableline
Mean\tablenotemark{a}\dotfill & 1.1 (0.5) & -360 (30) & -135 (40) & -19 (22) & 1.18 (0.41) & 10.9 (3.4) & 0.1 (0.1) & \nod & -440 (60) & 410 (70) & 11.5 (10.0) & 2.8 (3.6) & \nod \\
\cutinhead{CTTS: Medium 1~\micron\ Veiling}
DS Tau\dotfill   & 0.4\phm{*} & -400 & -130 & -42 & 0.58 & 4.2 & 0.2 & r & -500 & 400 & 2.0 & 1.4 & b,r  \\
HK Tau\dotfill   & 0.4\phm{*} & -350 & -128 & -39 & 0.32 & 2.7 & \nod & \nod & -450 & 300 & 1.1 & 1.0 & b,r  \\
UY Aur\dotfill   & 0.4\phm{*} & -350 & -105 & -26 & 0.60 & 4.2 & \nod & \nod & -300 & 300 & 4.5 & 0.4 & r  \\
                 & 0.4* & -350 & -59 & -7 & 0.57 & 3.3 & \nod & \nod & -300 & 300 & 3.6 & 0.8 & b,r  \\
YY Ori\dotfill   & 0.4\phm{*} & -350 & -122 & -48 & 0.60 & 3.6 & 1.1 & r & -350 & 450 & 1.4 & 3.6 & b,r  \\
BP Tau\dotfill   & 0.3\phm{*} & -350 & -94 & -19 & 0.58 & 3.6 & \nod & \nod & -250 & 250 & 5.8 & \nod & \nod  \\
DF Tau\dotfill   & 0.3\phm{*} & -300 & -76 & -12 & 0.71 & 4.2 & \nod & \nod & -200 & 300 & 3.6 & 1.3 & b  \\
DO Tau\dotfill   & 0.3\phm{*} & -300 & -64 & -9 & 1.35 & 7.8 & \nod & \nod & -200 & 300 & 2.4 & 2.5 & b  \\
GG Tau\dotfill   & 0.3\phm{*} & -300 & -126 & -18 & 0.40 & 3.3 & \nod & \nod & -350 & 450 & 5.1 & 0.2 & b  \\
GK Tau\dotfill   & 0.3\phm{*} & -300 & -109 & -4 & 0.24 & 1.8 & \nod & \nod & -300 & 350 & 2.1 & 2.8 & c  \\
GW Ori\dotfill   & 0.3\phm{*} & -400 & -115 & -2 & 0.50 & 4.7 & \nod & \nod & -450 & 400 & 5.0 & 4.4 & b  \\
UZ Tau E\dotfill & 0.3\phm{*} & -250 & -157 & -51 & 0.47 & 4.0 & \nod & \nod & -300 & 300 & 3.4 & 0.2 & r  \\
\tableline
Mean\tablenotemark{a}\dotfill & 0.3 (0.1) & -330 (50) & -110 (30) & -23 (19) & 0.57 (0.29) & 3.9 (1.5) & 0.1 (0.3) & \nod & -330 (100) & 350 (70) & 3.2 (1.6) & 1.7 (1.5) & \nod  \\
\cutinhead{CTTS: Low 1~\micron\ Veiling, Narrow \pgamma}
AA Tau\dotfill   & 0.2\phm{*} & -350 & -94 & -46 & 0.29 & 1.5 & 0.5 & r & -400 & 400 & 4.4 & 0.6 & r  \\
                 & 0.1* & -300 & -37 & 23 & 0.16 & 0.9 & \nod & \nod & -400 & 400 & 1.9 & 2.0 & r  \\
CY Tau\dotfill   & 0.1\phm{*} & -350 & -47 & -6 & 0.42 & 2.0 & \nod & \nod & -250 & 250 & 2.4 & 2.5 & b,r  \\
GI Tau\dotfill   & 0.1\phm{*} & -350 & -45 & -1 & 0.53 & 3.2 & 0.3 & r & -350 & 400 & 2.2 & 3.7 & b,r  \\
LkCa 8\dotfill   & 0.05\phm{*} & -300 & -34 & -4 & 0.16 & 1.0 & \nod & \nod & -250 & 350 & 1.7 & 1.5 & r  \\
DN Tau\dotfill   & 0.0\phm{*} & -400 & -28 & 4 & 0.24 & 1.1 & \nod & \nod & -350 & 350 & 2.1 & 1.7 & b,r  \\
DQ Tau\dotfill   & 0.0\phm{*} & -200 & -42 & 0 & 0.20 & 0.8 & \nod & \nod & -450 & 350 & 9.4 & 0.3 & b  \\
TW Hya\dotfill   & 0.0* & -350 & -58 & 0 & 1.71 & 8.8 & \nod & \nod & -350 & 350 & 13.6 & 5.3 & b  \\
                 & 0.0\phm{*} & -250 & -54 & 2 & 1.64 & 8.6 & \nod & \nod & -350 & 250 & 18.1 & 4.0 & b  \\
V836 Tau\dotfill & 0.0\phm{*} & -100 & -26 & 3 & 0.22 & 0.6 & \nod & \nod & -150 & 350 & 1.7 & 2.2 & b,r  \\
XZ Tau\dotfill   & 0.0\phm{*} & -300 & -44 & 1 & 0.21 & 1.0 & \nod & \nod & -325 & 250 & 2.8 & 1.2 & b  \\
\tableline
Mean\tablenotemark{a}\dotfill & 0.0 (0.0) & -290 (90) & -40 (10) & 2 (8) & 0.42 (0.50) & 2.2 (2.6) & 0.0 (0.1) & \nod & -320 (90) & 340 (50) & 4.2 (4.2) & 2.3 (1.5) & \nod  \\
\cutinhead{CTTS: Low 1~\micron\ Veiling, Broad \pgamma}
CI Tau\dotfill   & 0.2\phm{*} & \nod & \nod & \nod & 0.06 & 0.6 & \nod & \nod & -200 & 350 & 0.1 & 2.2 & b,r  \\
DE Tau\dotfill   & 0.2\phm{*} & -250 & -128 & -12 & 0.28 & 2.4 & \nod & \nod & -350 & 350 & 1.9 & 0.7 & b  \\
BM And\dotfill   & 0.1\phm{*} & -350 & -250 & -150 & 0.09 & 0.8 & 0.9 & r & -300 & 300 & 2.6 & 2.6 & r  \\
DD Tau\dotfill   & 0.1\phm{*} & -300 & -106 & -49 & 0.38 & 2.2 & \nod & \nod & -150 & 250 & 0.5 & 0.2 & b  \\
FP Tau\dotfill   & 0.1\phm{*} & -300 & -153 & -90 & 0.16 & 1.0 & \nod & \nod & -300 & 200 & 0.5 & 1.2 & b,r  \\
RW Aur B\dotfill & 0.1\phm{*} & -150 & -85 & -9 & 0.14 & 0.7 & 0.5 & r & -300 & 400 & 1.5 & 2.8 & r  \\
UZ Tau W\dotfill & 0.1\phm{*} & -300 & -91 & -44 & 0.25 & 1.4 & \nod & \nod & -250 & 250 & 0.3 & 0.8 & b,r  \\
SU Aur\dotfill   & 0.0\phm{*} & -400 & -199 & -120 & 0.17 & 1.4 & 0.6 & r & -450 & 200 & 0.1 & 2.1 & b,r  \\
UX Tau\dotfill   & 0.0\phm{*} & \nod & \nod & \nod & 0.09 & 1.3 & \nod & \nod & -400 & 300 & 2.3 & 2.0 & c  \\
\tableline
Mean\tablenotemark{a}\dotfill & 0.1 (0.1) & -280 (80) & -130 (40) & -54 (44) & 0.18 (0.11) & 1.3 (0.6) & 0.2 (0.3) & \nod & -300 (90) & 290 (70) & 1.1 (1.0) & 1.6 (0.9) & \nod  \\

\enddata
\tablecomments{Col 2: An asterisk following r$_Y$ identifies the observation chosen for the reference sample. Col 6:
Peak emission intensity in units of the continuum. Col 8,13: equivalent width of all absorption below the continuum.
Col. 9,14: Type of absorption (b) blue; (c) central; (r) red }

\tablenotetext{a}{Mean profile parameters for the reference sample from
each veiling group; standard deviation in parentheses.}
\end{deluxetable}

\clearpage
\begin{deluxetable}{lccccc}
\tablecaption{Blue \helium\ Absorption in 27 CTTS \label{t.bluabs}}
\tabletypesize{\scriptsize}
\tablewidth{0pt}
\tablehead{ & \colhead{$W_\lambda$} & \colhead{\%} & \colhead{$\Delta V$} &
\colhead{$V_{\rm{min}}$} &
\colhead{$V_{\rm{max}}$} \\
\colhead{Object} & \colhead{(\AA)} & \colhead{Abs.} & \colhead{(\kms)} & \colhead{(\kms)}
& \colhead{(\kms)}\\
\colhead{(1)} & \colhead{(2)} & \colhead{(3)} & \colhead{(4)} & \colhead{(5)} & \colhead{(6)}}
\startdata
AS 353 A\dotfill & 2.8 & 60 & 220 & -80 & -300  \\
CI Tau\dotfill   & 1.2 & 43 & 140 & -60 & -200  \\
CY Tau\dotfill   & 1.5 & 68 & 130 & -70 & -200  \\
DD Tau\dotfill   & 0.2 & 15 & 160\tablenotemark{a} & -10 & -170  \\
DE Tau\dotfill   & 0.7 & 16 & 190 & -140 & -330  \\
DF Tau\dotfill   & 1.3 & 49 & 150 & -50 & -200  \\
DG Tau\dotfill   & 1.2 & 33 & 200 & -250 & -450  \\
DK Tau\dotfill   & 3.7 & 80 & 410 & -40 & -450  \\
DL Tau\dotfill   & 0.4 & 28 & 60 & -100 & -160  \\
DN Tau\dotfill   & 0.3 & 13 & 120 & -210 & -330  \\
DO Tau\dotfill   & 2.5 & 69 & 160 & -40 & -200  \\
DQ Tau\dotfill   & 0.3 & 13 & 90 & -160 & -250  \\
DR Tau\dotfill   & 10.7 & 95 & 470 & 70 & -400  \\
DS Tau\dotfill   & 0.3 & 23 & 60 & -80 & -140  \\
FP Tau\dotfill   & 0.8 & 31 & 120 & 30\tablenotemark{b} & -90  \\
GG Tau\dotfill   & 0.2 & 15 & 60 & -150 & -210  \\
GI Tau\dotfill   & 0.6 & 15 & 180 & -140 & -320  \\
GW Ori\dotfill   & 4.4& 73 & 380\tablenotemark{a}  & -60 & -440  \\
HK Tau\dotfill   & 0.5 & 13 & 270 & -150 & -420  \\
HL Tau\dotfill   & 3.8 & 55 & 380 & -20 & -400  \\
SU Aur\dotfill  & 0.6 & 20 & 330\tablenotemark{a} & -90 & -420  \\
TW Hya\dotfill   & 5.3 & 98 & 230 & -100 & -330  \\
UY Aur\dotfill   & 0.4 & 45 & 60 & -90 & -150  \\
UZ Tau W\dotfill & 0.3 & 10 & 190 & -60 & -250  \\
V836 Tau\dotfill & 0.2 & 51 & 30 & 0 & -30  \\
XZ Tau\dotfill   & 1.2 & 46 & 220 & -100 & -320  \\
YY Ori\dotfill & 0.9 & 40 & 110 & 0\tablenotemark{b} & -110  \\
\enddata
\tablecomments{Col 3: Percentage of continuum absorbed at the
deepest point of the profile; Col 4: Range of velocities for blue absorption;
Col 5: Minimum velocity of blue absorption; Col 6: Maximum velocity of blue absorption.}
\tablenotetext{a}{Two distinct blueshifted absorption components.}
\tablenotetext{b}{Blueshifted absorption component is blended with a
redshifted absorption component; $V_{\rm{min}}$ is the local maximum
in the blended region.}
\end{deluxetable}
\clearpage

\begin{deluxetable}{lcccccccc}
\tablecaption{Class I Objects\label{t.class1sam}}
\tabletypesize{\footnotesize}
\setlength{\tabcolsep}{0.04in}
\tablewidth{0pt}
\tablehead{ \colhead{Object} & \colhead{Sp Type} & \colhead{Ref} & \colhead{HJD} &
\colhead{$\log \dot{M}_{\rm{acc}}$} &  \colhead{$\log\dot{M}_{\rm{wind}}$} & \colhead{Ref} &
\colhead{$\dot{M}_{\rm{wind}}/\dot{M}_{\rm{acc}}$} \\ \colhead{(1)} &
\colhead{(2)} & \colhead{(3)} & \colhead{(4)} & \colhead{(5)} &
\colhead{(6)} & \colhead{(7)} & \colhead{(8)}}
\startdata
SVS 13\dotfill     & K:   & 1 & 605.8,606.8 & -5.2\tablenotemark{a}  & -6.2 & 3 & 0.10\tablenotemark{a} \\
04248+2612\dotfill & M5.5 & 2 & 606.9 & -9.0  & -9.1 & 2 & 0.79 \\
04303+2240\dotfill & M0.5 & 2 & 605.9 & -6.1  & -6.8 & 2 & 0.20 \\
\enddata

\tablecomments{
Col 4: Heliocentric Julian Date (2,452,000 +); Col 5: mass accretion rate in \msunyr; Col 6:
mass loss rate in \msunyr; Col 8: Ratio of mass loss rate to mass accretion rate.}
\tablenotetext{a}{accretion rate assumed to be 10 times the mass-loss rate.}

\tablerefs{(1) \citealt{good86}; (2) \citealt{wh04}; (3) \citealt{dav01}.}
\end{deluxetable}

\clearpage

\begin{deluxetable}{lccccccccccccc}
\tablecaption{Class I Veiling and Profiles \label{t.class1kin}}
\rotate
\tabletypesize{\scriptsize}
\setlength{\tabcolsep}{0.04in}
\tablewidth{0pt}
\tablehead{\multicolumn{2}{c}{} & \multicolumn{7}{c}{\pgamma} &
\multicolumn{5}{c}{He I $\lambda$10830} \\ \multicolumn{2}{c}{} &
\multicolumn{7}{c}{------------------------------------------------------------------------------------------}
&
\multicolumn{5}{c}{-----------------------------------------------------------------}\\
& & \colhead{$V_{\rm{bwing}}$} & \colhead{\vbhi} & \colhead{\vc} &
\colhead{Peak} & \colhead{Em $W_\lambda$} & \colhead{Ab $W_\lambda$} &
\colhead{Ab} & \colhead{$V_{\rm{bwing}}$} & \colhead{$V_{\rm{rwing}}$}
& \colhead{Em $W_\lambda$} & \colhead{Ab $W_\lambda$} & \colhead{Ab}
\\ \colhead{Object} & \colhead{$r_Y$} & \colhead{(\kms)} &
\colhead{(\kms)} & \colhead{(\kms)} &
\colhead{Intensity\tablenotemark{a}} & \colhead{(\AA)} &
\colhead{(\AA)} & \colhead{Type\tablenotemark{b}} & \colhead{(\kms)} &
\colhead{(\kms)} & \colhead{(\AA)} & \colhead{(\AA)} &
\colhead{Type\tablenotemark{b}}}
\startdata
SVS 13\dotfill & 0.4\phm{*} & -200 & -79 & 15 & 0.77 & 5.4 & \nod & \nod & -350 & 100 & \nod & 4.7 & b \\
               & 0.3 & -200 & -58 & 25 & 0.69 & 4.2 & \nod & \nod & -350 & 100 & \nod & 4.4 & b \\
04248+2612\dotfill & 0.0\phm{*} & -200 & -45 & -4 & 0.41 & 1.7 & \nod & \nod & -350 & 200 & 3.5 & 0.7 & b \\
04303+2240\dotfill & 1.8\phm{*} & -350 & -111 & -11 & 2.08 & 15.4 & \nod & \nod & -350 & 300 & 12.3 & \nod & \nod \\
\enddata

\tablenotetext{a}{Peak emission intensity in units of the continuum.}
\tablenotetext{b}{Classification of absorption components: (b) blue.}
\end{deluxetable}

\clearpage

\begin{deluxetable}{lcccccccc}
\tablecaption{CTTS Mean B, V, Y, K Veiling \label{t.veiling}}
\tabletypesize{\scriptsize}
\tablewidth{0pt}
\tablehead{\colhead{Object} & \colhead{$r_B$} & \colhead{$r_V$}
& \colhead{$r_Y$} & \colhead{$r_K$} & \colhead{$r_B/r_Y$} & \colhead{$r_V/r_Y$} &
\colhead{$r_K/r_Y$}}
\startdata
AA Tau\dotfill   & 0.27 & 0.31  & 0.15  & 0.38 & 1.80 & 2.07 & 2.53 \\
AS 353 A\dotfill & 3.45 & 5.10  & 1.90  & \nod & 1.82 & 2.68 & \nod \\
BM And\dotfill   & \nod & \nod  & 0.10  & 1.40 & \nod & \nod & 14.00 \\
BP Tau\dotfill   & 0.80 & 0.71  & 0.30  & 0.72 & 2.67 & 2.37 & 2.40 \\
CI Tau\dotfill   & \nod & 0.47  & 0.20  & \nod & \nod & 2.35 & \nod \\
CW Tau\dotfill   & \nod & 1.25  & 1.15  & \nod & \nod & 1.09 & \nod \\
CY Tau\dotfill   & 0.82 & 1.20  & 0.10  & \nod & 8.20 & 12.00 & \nod \\
DD Tau\dotfill   & \nod & 2.90  & 0.10  & \nod & \nod & 29.00 & \nod \\
DE Tau\dotfill   & 0.88 & 0.57  & 0.20  & 1.05 & 4.40 & 2.85 & 5.25 \\
DF Tau\dotfill   & 1.25 & 0.93  & 0.30  & 0.81 & 4.17 & 3.10 & 2.70 \\
DG Tau\dotfill   & 2.70 & 3.15  & 0.80  & 1.43 & 3.38 & 3.94 & 1.79 \\
DK Tau\dotfill   & 0.69 & 0.49  & 0.50  & 1.62 & 1.38 & 0.98 & 3.24 \\
DL Tau\dotfill   & 2.64 & 1.64  & 1.10  & \nod & 2.40 & 1.49 & \nod \\
DN Tau\dotfill   & 0.16 & 0.07  & 0.00  & 0.21 & 6.40 & 2.80 & 8.40 \\
DO Tau\dotfill   & \nod & 4.70  & 0.30  & 1.90 & \nod & 15.67 & 6.33 \\
DQ Tau\dotfill   & \nod & 0.18  & 0.00  & 0.30 & \nod & 7.20 & 12.00 \\
DR Tau\dotfill   & 4.30 & 8.14  & 2.00  & 4.00 & 2.15 & 4.07 & 2.00 \\
DS Tau\dotfill   & 1.00 & 0.83  & 0.40  & 1.10 & 2.50 & 2.07 & 2.75 \\
FP Tau\dotfill   & \nod & 0.14  & 0.10  & 0.40 & \nod & 1.40 & 4.00 \\
GG Tau\dotfill   & 0.90 & 0.30  & 0.30  & 0.26 & 3.00 & 1.00 & 0.87 \\
GI Tau\dotfill   & 0.60 & 0.24  & 0.10  & 0.72 & 6.00 & 2.40 & 7.20 \\
GK Tau\dotfill   & 0.11 & 0.17  & 0.30  & 1.11 & 0.37 & 0.57 & 3.70 \\
GM Aur\dotfill   & 0.19 & 0.22  & 0.00  & 0.30 & 7.60 & 8.80 & 12.00 \\
GW Ori\dotfill   & 0.00 & \nod  & 0.30  & \nod & 0.00 & \nod & \nod \\
HK Tau\dotfill   & \nod & 1.10  & 0.40  & \nod & \nod & 2.75 & \nod \\
HL Tau\dotfill   & \nod & \nod  & 1.00  & \nod & \nod & \nod & \nod \\
HN Tau\dotfill   & 0.60 & 0.76  & 0.50  & \nod & 1.20 & 1.52 & \nod \\
LkCa 8\dotfill   & 0.15 & 0.15  & 0.05  & \nod & 3.00 & 3.00 & \nod \\
RW Aur A\dotfill & 3.20 & 2.04  & 0.90  & \nod & 3.56 & 2.27 & \nod \\
RW Aur B\dotfill & \nod & \nod  & 0.10  & \nod & \nod & \nod & \nod \\
SU Aur\dotfill   & \nod & \nod  & 0.00 & 0.60 & \nod & \nod & 24.00 \\
TW Hya\dotfill   & \nod & 0.29  & 0.00 & 0.09 & \nod & 11.60 & 3.60 \\
UX Tau\dotfill   & \nod & \nod  & 0.00 & \nod & \nod & \nod & \nod \\
UY Aur\dotfill   & 0.75 & 0.53  & 0.40 & 1.65 & 1.88 & 1.32 & 4.12 \\
UZ Tau E\dotfill & \nod & 0.73  & 0.30 & \nod & \nod & 2.43 & \nod \\
UZ Tau W\dotfill & \nod & \nod  & 0.10 & \nod & \nod & \nod & \nod \\
V836 Tau\dotfill & \nod & 0.04  & 0.00 & \nod & \nod & 1.60 & \nod \\
XZ Tau\dotfill   & \nod & \nod  & 0.00 & \nod & \nod & \nod & \nod \\
YY Ori\dotfill   & \nod & 1.80  & 0.40 & \nod & \nod & 4.50 & \nod \\

\enddata

\tablecomments {1. Veilings are means from the following compilations:
$r_B$: \citet{bb90}, \citet{GHBC}; $r_V$: \citet{ab02}, \citetalias{HEG}, \citet{mhc98b};
$r_Y$: this paper;
$r_K$: \citet{folha99}, \citet{jkv01},\citet{muz03} 2. In calculating ratios, objects with $r_Y$ non detections are set
to $r_Y=0.025$.  }
\end{deluxetable}

\clearpage

\epsscale{0.9}
\clearpage

\epsscale{1.0}

\clearpage

\epsscale{0.9}

\clearpage

\clearpage
\clearpage

\clearpage

\epsscale{1.0}
\clearpage

\epsscale{0.8}
\clearpage

\epsscale{1.0}
\clearpage

\clearpage

\clearpage

\clearpage

\clearpage

\clearpage

\end{document}